\documentclass{article}
\usepackage{arxiv}
 
 \usepackage{hyperref}
 \usepackage{color}
 \usepackage{bm}
 \usepackage{graphicx, subcaption}
 \usepackage{latexsym}
 \usepackage{pifont}
 \usepackage{amssymb}
 \usepackage{hyperref}
 \usepackage{titlesec}
 \expandafter\let\csname equation*\endcsname\relax
 \expandafter\let\csname endequation*\endcsname\relax
 \usepackage{amsmath}
 \usepackage{listings}
 \usepackage{verbatim}
 \usepackage{geometry}
 \usepackage{booktabs}
 \usepackage{enumitem}
 \usepackage[normalem]{ulem} 
 \usepackage{breqn}
 \usepackage[toc,page]{appendix}
 \usepackage{cite}
 
 \definecolor{red}{rgb}{1.0,0.0,0.0}
 \definecolor{gre}{rgb}{0.0,1.0,0.0}
 \definecolor{blu}{rgb}{0.0,0.0,1.0}

 \definecolor{ora}{rgb}{1.0,0.5,0.0}
 \definecolor{gra}{rgb}{0.0,0.5,1.0}

 \newcommand{\change}[2]{#2}

 \newcommand{\changenew}[2]{#2}
 
 \newcommand{\ba}{\begin{abstract}}
 \newcommand{\ea}{\end{abstract}}
 \newcommand{\be}{\begin{equation}}
 \newcommand{\ee}{\end{equation}}

 \newcommand{\ie}{\textit{i}.\textit{e}., }
 \newcommand{\eg}{\textit{e}.\textit{g}. }

 \newcommand{\rom}[1]{\uppercase\expandafter{\romannumeral #1 \relax}}
 \newcommand{\Poincare}{Poincar$\acute{\rm e}$ }
 
 \newcommand{\Fig}[1]{figure \ref{fig:#1}}
 
 \newcommand{\Eqn}[1]{equation (\ref{eq:#1})}
 
 \renewcommand{\t}{\theta}
 \newcommand{\z}{\zeta} 
 \newcommand{\ds}{\displaystyle}
 \newcommand{\vect}[1]{\mathbf{#1}}
 \newcommand{\dd}[1]{\textnormal{d} #1}
 \newcommand{\pdv}[2]{\frac{\partial #1}{\partial #2}}
 
 \newcommand{\hdv}[3]{\frac{\partial^2 #1}{\partial #2 \partial #3}}
 
 \newcommand{\abs}[1]{|#1|}
 \newcommand{\grad}{\nabla}
 \newcommand{\tento}[1]{\times 10^{#1}}
 
 \newcommand{\half}{\frac{1}{2}}
 
\graphicspath{{./plots/}}
\usepackage[utf8]{inputenc}
\title{Identification of \change{dangerous}{important} error fields in stellarators using Hessian matrix method}
\author[1*]{Caoxiang Zhu}
\author[1]{David A. Gates}
\author[1]{Stuart R. Hudson}
\author[2]{Haifeng Liu}
\author[2]{Yuhong Xu}
\author[3]{Akihiro Shimizu}
\author[3]{Shoichi Okamura}

\affil[1]{Princeton Plasma Physics Laboratory, P.O. Box 451, New Jersey 08543-0451, USA}
\affil[2]{Institute of Fusion Science, Southwest Jiaotong University, Chengdu 610031, P. R. China}
\affil[3]{National Institute for Fusion Science, National Institutes of Natural Sciences, Toki 509-5292, Japan}
\affil[*]{\textit{Email: czhu@pppl.gov}}
\renewcommand\Authands{ and }
\begin{document}
\maketitle
\ba
Error fields are predominantly attributed to inevitable coil imperfections. Controlling error fields during coil fabrication and assembly is crucial for stellarators. Excessively tight coil tolerance increases time and cost, and,  in part, led to the cancellation of NCSX and delay of W7-X. In this paper, we improve the recently proposed Hessian matrix method to rapidly identify \change{dangerous}{important} coil deviations. Two of the most common figures of merit, magnetic island size and quasi-symmetry, are analytically differentiated over coil parameters. By extracting the eigenvectors of the Hessian matrix, we can directly identify sensitive coil deviations in the order of the eigenvalues. The new method is applied to the upcoming CFQS configuration. \change{Dangerous}{Important} perturbations that enlarge n/m=4/11 islands and deteriorate quasi-axisymmetry of the magnetic field are successfully determined. The results suggest each modular coil should have separate tolerance and some certain perturbation combinations will produce significant error fields. By relaxing unnecessary coil tolerance, this method will hopefully lead to substantial a reduction in time and cost.
\ea
\section{Introduction}
Plasma performance in magnetically confined fusion devices depends on the quality of \changenew{}{the} magnetic field. 
Magnetic field irregularities, namely `error fields', can lead to \changenew{}{the} destruction of magnetic surfaces \cite{Rosenbluth1966} and locked modes \cite{LaHaye1992}. 
One of the main sources of error field is inevitable coil deviations.
The stellarator generally has more complicated coils than axisymmetric devices (like tokamaks) and the confining magnetic field \changenew{predominately}{predominantly} arises from carefully shaped external coils.
Therefore, controlling error fields during coil fabrication and assembly is crucial for stellarator construction.
This is even more challenging than expected, as high accuracy requirement on modular coils was the largest driver of cost growth for the National Compact Stellarator Experiment (NCSX) and partly led to the project cancellation \cite{Neilson2010, Strykowsky2009}.
The high demand \changenew{of}{for} coil tolerance was also identified during the assembly of W7-X \cite{Bosch2017}.

As-built coil geometries might differ from designed models in location, orientation or even shape. 
Coil deviations, as measured between designed models and final built shapes, have different effects on plasma performance.
The cost and complexity of a device can increase dramatically with tight construction tolerance.
Thus, it is necessary to carry out \changenew{}{an} error field sensitivity analysis for identifying \change{dangerous}{important} coil deviations and determining acceptable tolerance prior to machine construction.

The Large Helical Devices (LHD) evaluated the irregular magnetic fields produced by global deformations and local irregularities on coils.
Poloidal field (PF) and helical coils were represented with Fourier coefficients and the vacuum flux surfaces were plotted out when perturbing different Fourier modes.
It was found that horizontal shifts ($n=1$) had the most significant influence on \changenew{}{the} destruction of magnetic surfaces \cite{Yamazaki1993}. 
On the Columbia Non-neutral Torus (CNT), Kremer \cite{Kremer_thesis} analyzed the volume of magnetic surfaces under hundreds of random coil displacements, including shifts and rotations of coils and he suggested the PF coils could be misplaced by a maximum distance of 1 cm while inter-linked (IL) coils could only be perturbed up to 2 mm.
This was also observed later by Hammond \etal \cite{Hammond2016} when the gradient of the rotational transform on the magnetic axis with respect to several defined rigid displacements were calculated by the finite difference.
NCSX investigated \changenew{}{the} impacts of systematic coil geometric perturbations and tolerance schemes on magnetic island size \cite{Brooks2003}.
Local errors including coil-plasma spacing, short wavelet orthogonal displacements, and broad coil deformations causing coil length errors were also examined \cite{Williamson2005}.
The results from NCSX suggested that modular coils required more tight tolerance (about 1.5 mm) than PF \& TF coils, and particularly the inboard regions of modular coils had more significant effects on flux quality, while errors in other regions might approach 3 mm or even larger coil tolerance.
As the largest stellarator, the Wendelstein 7-X (W7-X) performed extensive studies on coil tolerance \cite{Lazerson2018}.
The primary criteria \changenew{was}{were} the resonant magnetic perturbations, $B_{11}, B_{22}, B_{33} \ \& \ B_{44}$ since W7-X has a $m/n=5/5$ island chain outside the last closed flux surface (LCFS).
Numerical investigations \cite{Rummel2004, Kisslinger2005} showed that the resonant magnetic field perturbations were the most sensitive to rotations of coils and modules, after the effects of manufacturing errors, shifts and rotations of individual coils and modules were compared.
With numerous efforts and advanced manufacturing techniques, W7-X superconducting coil system was built and assembled with impressively high accuracy.
The maximum deviation for non-planar coils from the average shape is of the order of 2 mm \cite{Andreeva2009} and the average alignment deviation for all 70 main field coils after the assembly is 1.2 mm \cite{Bosch2017}.

More recently, there are several new approaches proposed to address the challenge of tight coil tolerance for stellarators. 
A shape gradient method was developed by Landreman \& \change{Elizabeth}{Paul} \cite{Landreman2018} to compute local coil tolerance after calculating all the derivatives of a figure of merit with respect to coil parameters.
Coil shapes with respect to plasma boundary are analytically differentiated by Hudson \etal \cite{Hudson2018} and it can be used to find simpler coils.
Lobsien \etal \cite{Lobsien2018} adopted \changenew{}{a} stochastic optimization method to find more robust coils with higher tolerance could by evaluating numerous randomized coil perturbations using a Monte Carlo sampling approach in optimization runs.

Most of the error field sensitivity studies to date are carried out by perturbing (shifting or rotating) coils in certain potential directions and then evaluate the changes in the figure of merit, which is normally the quality of magnetic surfaces, \ie magnetic islands, as the resonant magnetic perturbation will be amplified, destruct the flux surfaces and degrade plasma confinement.
This approach has been successfully applied to several devices.
However, it requires massive computation resources, perhaps as well as man-hours, to scan all possible individual coil deviations and compare different combination scenarios.
In this paper, we propose a new method to rapidly identify \change{dangerous}{important} coil deviations that could possibly appear during coil fabrication and assembly.

The Hessian matrix can be used for sensitivity analysis \cite{Boozer2015Design}.
In \cite{Zhu_2018_Hessian}, a Hessian matrix method was described to determine error field sensitivity to coil deviations.
The figure of merit used was the root-mean-squared (RMS) normal field error on the target plasma boundary, which comes from the coil design code FOCUS \cite{Zhu_2017_FOCUS}. 
A quadratic approximation indicates coil perturbation in the direction of \changenew{}{the} eigenvector corresponding to the largest eigenvalue has the most significant effect on the figure of merit.
The Hessian matrix method was then applied to a CNT-like configuration as a proof of principle and the results were consistent with previous observations.
The RMS normal field error represents the discrepancy between the desired magnetic field and the one produced by coils, but it does not have any particular physics meaning.
In practical situations, what should be considered when evaluating the error field is \changenew{the}{} plasma confinement performance.
In this paper, we will implement the Hessian matrix method over the magnetic island size, which is used in most error field studies.
Additionally, we would also evaluate the so-called `quasi-symmetry' of \changenew{}{the} magnetic field \cite{Nuhrenberg1988}, which is the symmetry of magnetic field strength $|\vect{B}|$.
Quasi-symmetry has been both theoretically predicted and experimentally confirmed to reduce neoclassical transport and is one of the key qualities used in today's stellarator optimizations.

This paper is organized as follows.
In Sec. \ref{method}, we have a brief view of the Hessian matrix method.
The two new figures of merit, island size and quasi-symmetry, are also described in Sec. \ref{method}.
In Sec. \ref{application}, we apply the method to an actual device, the Chinese First Quasi-axisymmetric Stellarator (CFQS) \cite{xu2019}, which is currently under construction.
\changenew{}{The} results of analyzing the most \change{dangerous}{important} islands and evaluations on quasi-axisymmetry (QA) are presented.
A method for computing the maximum allowable coil deviation under the worst scheme is discussed in Sec. \ref{result_tol} and a demonstration using the information from \changenew{}{the} analysis of magnetic islands and quasi-symmetry to improve coil designs is shown in Sec. \ref{QA_opt}.
We will conclude in Sec. \ref{discussion}.

\section{Hessian matrix method, magnetic island size and quasi-symmetry} \label{method}
\subsection{Hessian matrix method}
Coil optimization consists of varying coil parameters, in whatever representation, and minimizing an objective function which is the weighted summation of multiple penalty functions.
Once optimal coils are found, a small change in coils shapes (and currents), which can \changenew{}{be} described as $\delta \vect{x}$ in parameter space, will cause a departure in the figure of merit away from the optimum. 
This change can be approximated by
\be \label{eq:quadratic}
\delta F \approx \half {\delta \vect{x}}^{\mathrm{T}} \cdot \vect{H}_0 \cdot {\delta \vect{x}} \ ,
\ee
where the matrix $\vect{H}_0$ is known as the Hessian matrix (second-order derivatives).
\change{}{In \Eqn{quadratic}, only the quadratic term is left since the linear term is zero at the optimum (stationary point) and we are only considering a small perturbation (neglecting higher-order terms).}
The Hessian matrix is symmetric and its eigenvalues are positive.
By perturbing coils in the direction of eigenvector, $\delta \vect{x} = \xi \vect{v}_i $, the change of the figure of merit is 
\be
\delta F \approx \half \lambda_i \xi^2 \ .
\ee
Its eigenvectors can be ordered by the magnitude of associated eigenvalues $\lambda_i$.
The first \changenew{principle}{principal} eigenvector, which is the one corresponding to the largest eigenvalue, describes the most sensitive perturbation.
With the information in eigenvectors and eigenvalues of the Hessian matrix, we can easily identify the sensitivity of \changenew{}{the} error field to coil deviations.
More details about the Hessian matrix method can be found in \cite{Zhu_2018_Hessian}.

\subsection{Magnetic island size}
For magnetic fields with a continuously nested family of flux surfaces, small perturbations may break the surfaces.
The destroyed magnetic surface has \changenew{}{a} rational rotational transform ($\iota=n/m$) which is in resonance with perturbations.
Magnetic islands are then formed.
Plasma confinement across islands is different from within magnetic surfaces and magnetic islands can give rise to plasma instabilities \cite{Waelbroeck2009}.
For this reason, an important figure of merit to measure the quality of \changenew{}{the} magnetic field is the size of islands, especially in the core.

The width of a magnetic island in toroidal flux space is calculated as \cite{Boozer2015Review}
\be \label{eq:island_width}
w = 4 \sqrt{\frac{b_{mn}}{m \iota'_{mn}}} \ ,
\ee
where $m$ the poloidal mode number, and $\iota'_{mn}=\dd{\iota}/\dd{\psi_t}|_{\iota=n/m}$.
\change{}{The resonant Fourier component, $b_{mn}$, is decomposed at the associated rational surface,
\be \label{eq:bmn}
b_{mn} = \frac{1}{2 \pi^2}\int_0^{2\pi} \int_0^{2\pi} \frac{\vect{B} \cdot \grad{\psi_t}}{\vect{B}_0 \cdot \grad{\zeta}} \  e^{-i(m\t - n\z)} \dd{\t} \dd{\z} \ .
\ee
In this calculation, we write the magnetic field as $\vect{B} = \vect{B_0} + \delta \vect{B}$, where $\vect{B_0}$ is the nearby magnetic field with perfect magnetic surfaces and $\delta \vect{B}$ is a small perturbation.}
Island width is proportional to the square root of the magnitude of $b_{mn}$, therefore even a small perturbation can generate sizable islands.
This is particularly true for low poloidal number islands in low shear configurations. 

For a given configuration, one could find the most \change{dangerous}{important} island chains that the rotational transform profile will cross.
For instance, NCSX cares $\iota=3/7$, $3/6$ and $3/5$ \cite{Hudson2002}.
Furthermore, some configurations might have special requirements on finite-size islands, \eg W7-X has $\iota=5/5$ island divertors \cite{Strumberger1996}.
To manipulate islands, pre-selected resonant components should be maintained to target values, which is to minimize the following functional,
\be \label{eq:frp}
F_{RP} = \sum_{m,n} (b_{mn} - b^o_{mn})^2 \ ,
\ee
where $b^o_{mn}$ is the target value of $b_{mn}$ \change{}{and the summation is over the chosen resonant components}.
To suppress island, $b^o_{mn}$ should be zero.
\change{}{Here, we have considered only the effect of the resonant perturbation in \Eqn{island_width} since the rotational transform profile will not change substantially under small perturbations.}
\change{Here the summation is over the chosen resonant components.
Resonant Fourier component is decomposed at the associated rational surface,
$
b_{mn} = \frac{1}{2 \pi^2}\int_0^{2\pi} \int_0^{2\pi} \frac{\vect{B} \cdot \grad{\psi_t}}{\vect{B}_0 \cdot \grad{\zeta}} \  e^{-i(m\t - n\z)} \dd{\t} \dd{\z} \ ,
$
where $\vect{B_0}$ is the ideal magnetic field and has perfect magnetic surfaces, $\vect{B} = \vect{B_0} + \delta \vect{B}$ the perturbed magnetic field, $(\psi_t, \t, \z)$ the straight field line coordinates.}{}

Accurately computing the derivatives of $b_{mn}$ with respect to coil parameters is not easy.
\change{One of the reasons is that the position and shape of rational surfaces depend on the magnetic field.
The integration in \Eqn{bmn} changes when coil shapes are varied.
For numerical simplicity, we will assume that the rational surface does not move or deform under \emph{small} perturbations.
Hence, only the magnetic field $\vect{B}$ will change in \Eqn{bmn}.}
{As the magnetic field is changed under coil perturbations, the resonant harmonics, $b_{mn}$ in \Eqn{bmn}, varies in several ways: the perturbation field is changed, the flux surface is moved, the straight field line coordinates \changenew{is}{are} reconstructed and the nearby perfect magnetic field is altered.
In this paper, we shall only consider \emph{small} perturbations, by which we assume $\vect{B_0}$ is not altered and omit the motion of flux surfaces.
Hence, only the change in the perturbation field $\delta \vect{B}$ will be considered under small coil deformations.}
\changenew{}{This linear approximation is used in W7-X error field studies \cite{Lazerson2018, Rummel2004, Kisslinger2005, Andreeva2009, Bozhenkov2016} and on NCSX \cite{Brooks2003}.}
Rational surface information, the straight field line coordinates $(\psi_t, \t, \z)$ and the normal vector $\grad{\psi_t}$, could be obtained by the so-called ``quadratic-flux-minimizing surface'' \cite{Hudson1998} or approximated by a closest flux surface in free-boundary MHD equilibrium calculations (like VMEC \cite{Hirshman1986}).
Derivatives of $b_{mn}$ are now calculated as
\begin{align}
\ds \pdv{b_{mn}}{X_i} & = \frac{1}{2 \pi^2} \int_0^{2\pi} \int_0^{2\pi} \pdv{\vect{B}}{X_i} \cdot \frac{\grad{\psi_t}}{\vect{B}_0 \cdot \grad{\zeta}} \ e^{-i(m\t - n\z)} \dd{\t} \dd{\z} \ , \label{eq:frp_deriv1} \\ 
\ds \frac{\partial^2 b_{mn}}{\partial X_i \partial X_j} & = \frac{1}{2 \pi^2} \int_0^{2\pi} \int_0^{2\pi} \frac{\partial^2 \vect{B}}{\partial X_i \partial X_j} \cdot \frac{\grad{\psi_t}}{\vect{B}_0 \cdot \grad{\zeta}} \  e^{-i(m\t - n\z)} \dd{\t} \dd{\z}  \ , \label{eq:frp_deriv2}
\end{align}
where $X_i$ and $X_j$ are arbitrary terms in the coil parameter vector $\vect{X}$.
\changenew{}{In \Eqn{frp_deriv1} and \Eqn{frp_deriv2}, we are using the linear approximation and have considered only the changes in the perturbation field.}
Derivatives of $F_{RP}$ could be then obtained.
The first- and second-order functional derivatives of the magnetic field produced by external coils are calculated in FOCUS (Eq. (A.4) \& (A.5) in \cite{Zhu_2018_Hessian}).
We can then use this information to rapidly compute the Hessian matrix of the magnetic island metric.

\subsection{Quasi-symmetry}
The guiding center motion of collisionless particles was found to be governed by the magnetic field strength $B$ alone \cite{Boozer1983} in Boozer coordinates.
It implies that configurations with \changenew{a}{} symmetry in $B$ will have good neoclassical transport, although the magnetic field itself is not symmetric. 
The condition that $B$ only depends on one helicity is called `quasi-symmetry'.
This leads to an important category of optimized stellarator designs, like quasi-axisymmetric stellarators \cite{Garabedian1998, NCSX, HennebergQA}, and quasi-helical stellarators \cite{Nuhrenberg1988, HSX, Ku2011}. 

The quality of quasi-symmetry (on one flux surface) can be evaluated using
\be \label{eq:fqs}
\ds F_{QS} = \sum_{n/m \neq N/M} \left (\frac{B_{m,n}}{B_{0,0}} \right )^2 \ ,
\ee
where $N, M$ are the target Fourier modes to be conserved, \eg $N=0$ for quasi-axisymmetry (QA).
Note that $F_{QS} =0$ indicates perfect quasi-symmetry of the magnetic field on the reference flux surface.
The Fourier component of magnetic field strength, $B_{m,n}$, is then decomposed in Boozer angles $(\t_B, \z_B)$,
\be \label{eq:Bmn}
B_{m,n} = \frac{1}{2 \pi^2} \int_0^{2\pi} \int_0^{2\pi} \sqrt{\vect{B} \cdot \vect{B}} \  e^{-i(m\t_B - n N_p \z_B)} \dd{\t_B} \dd{\z_B} \ .
\ee
Here, $N_p$ is the number of periodicity.
\changenew{Again, if we assume the reference flux surface \change{does not move}{and the Boozer coordinates do not change} under \emph{small} perturbations, $B_{m,n}$ are likewise depending on $\vect{B}$ alone.}
{$B_{m,n}$ depends on the magnetic field $\vect{B}$, the flux surface and the Boozer coordinates on the surface.
When calculating the derivatives of $B_{m,n}$, a finite difference method can be used.
Each evaluation involves preparing the perturbed magnetic field, solving the free-boundary MHD equilibrium, constructing Boozer coordinates and calculating $B_{m,n}$.
To calculate the Hessian matrix, the finite difference method might be computationally expensive especially when the dimension of the Hessian matrix is large.
Instead of using the finite difference, in this paper we will apply the same linear approximation as the magnetic island to quickly construct the Hessian matrix.
}

\changenew{As in \Eqn{frp_deriv1} and \Eqn{frp_deriv2}, $B_{m,n}$ derivatives are calculated (they are not written down here for brevity),}
{If we apply the above linear approximation and consider only the change in the magnetic field $\vect{B}$, $B_{m,n}$ derivatives can be calculated as in \Eqn{frp_deriv1} and \Eqn{frp_deriv2} (they are not written down here for brevity).}
Consequently, the derivatives of $F_{QS}$ are computed as, 
\be
\ds \pdv{F_{QS}}{X_i} = \sum_{n/m \neq N/M} 2 \frac{B_{m,n}}{{B_{0,0}}^2} \left [ \pdv{B_{m,n}}{X_i} - \pdv{B_{0,0}}{X_i} \frac{B_{m,n}}{B_{0,0}}\right ] \ ,
\ee
\begin{align}
\ds \frac{\partial^2 F_{QS}}{\partial X_i \partial X_j}  = \sum_{n/m \neq N/M} \frac{2}{{B_{0,0}}^2} \bigg[ & \pdv{B_{m,n}}{X_i} \pdv{B_{m,n}}{X_j} + \frac{3{B_{m,n}}^2}{{B_{0,0}}^2} \pdv{B_{0,0}}{X_i} \pdv{B_{0,0}}{X_j}  \nonumber \\
    & + B_{m,n} \hdv{B_{m,n}}{X_i}{X_j} - \frac{{B_{m,n}}^2}{{B_{0,0}}} \hdv{B_{0,0}}{X_i}{X_j} \nonumber \\
    & - \frac{2{B_{m,n}}}{{B_{0,0}}} \left ( \pdv{B_{m,n}}{X_i} \pdv{B_{0,0}}{X_j} + \pdv{B_{0,0}}{X_i} \pdv{B_{m,n}}{X_j}  \right ) \bigg] \ .
\end{align}
These equations are evaluated in FOCUS and the Hessian matrix of the quasi-symmetry metric could be computed.

\section{Numerical application on CFQS} \label{application}
\subsection{Brief introduction of CFQS and its coils}
\begin{figure}[ht]
    \centering
    \includegraphics[width=.8\textwidth]{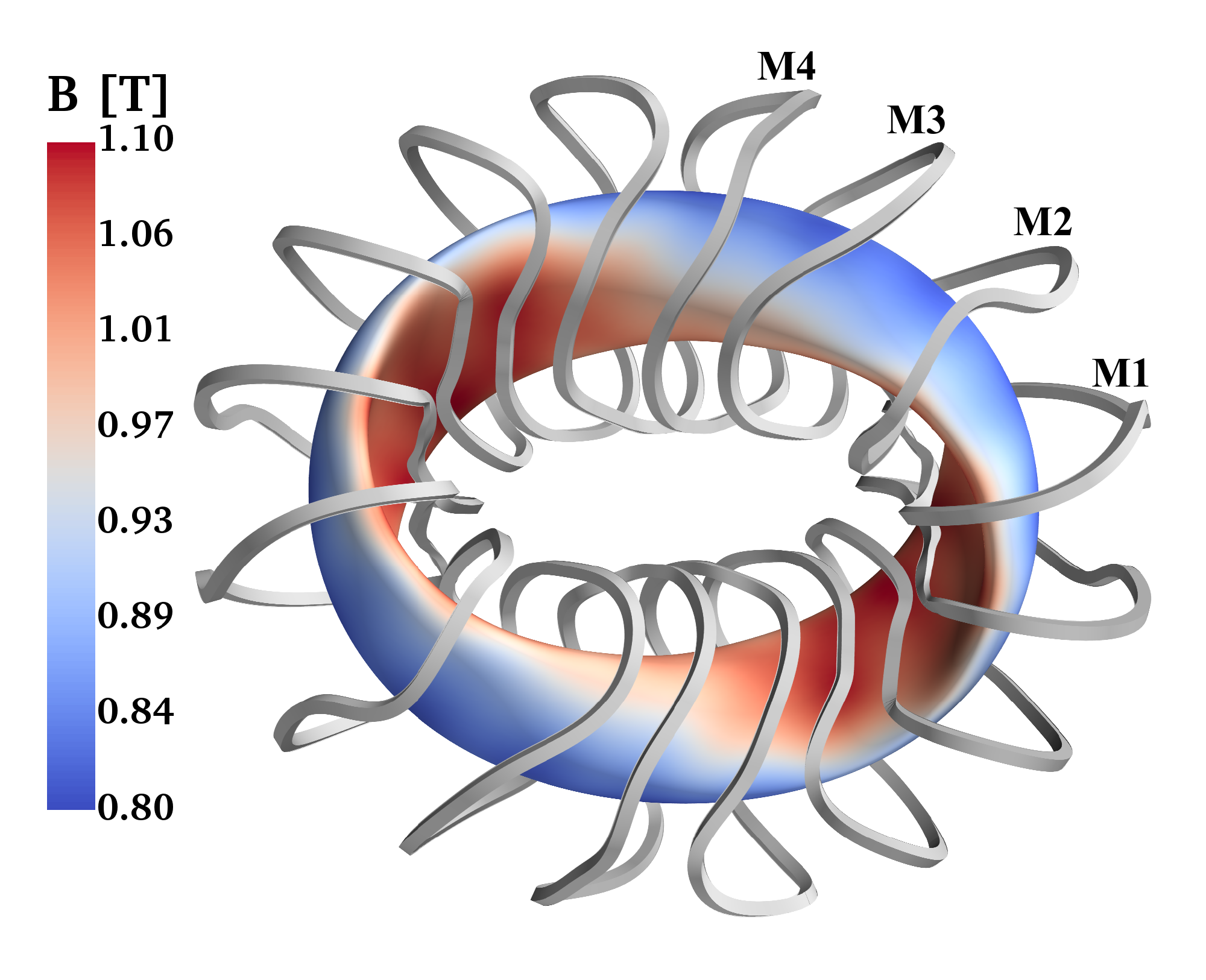}
    \caption{CFQS target boundary shape and modular coils. Colors on the boundary indicate the magnetic field strength produced by modular coils. Coils are shown with finite width for illustration not reflecting the actual engineering designs. }
    \label{fig:CFQS}
\end{figure}

CFQS is a quasi-axisymmetric stellarator being built in China under the collaboration of Southwest Jiaotong University in China and \changenew{}{the} National Institute for Fusion Science in Japan.
The main parameters of CFQS are as follows: the toroidal period number $N_p = 2$, major radius $R_0 =1.0$ m, minor radius $a =0.25$ m and magnetic field strength $B_t =1.0$ T \cite{xu2019}.
Sixteen modular coils are designed to provide a relatively accurate magnetic field along with adequate coil intervals and acceptable coil curvature \cite{Liu2018}.
The target plasma boundary of CFQS and modular coils are shown in \Fig{CFQS}.
Because of stellarator symmetry and periodicity, there are only four unique coil shapes, denoted as M1, M2, M3 and M4.
Among the four coil types, the M4 is considered to be the most complex one, as it has the largest toroidal excursions. 

\begin{figure}[ht]
    \centering
    \includegraphics[width=.8\textwidth]{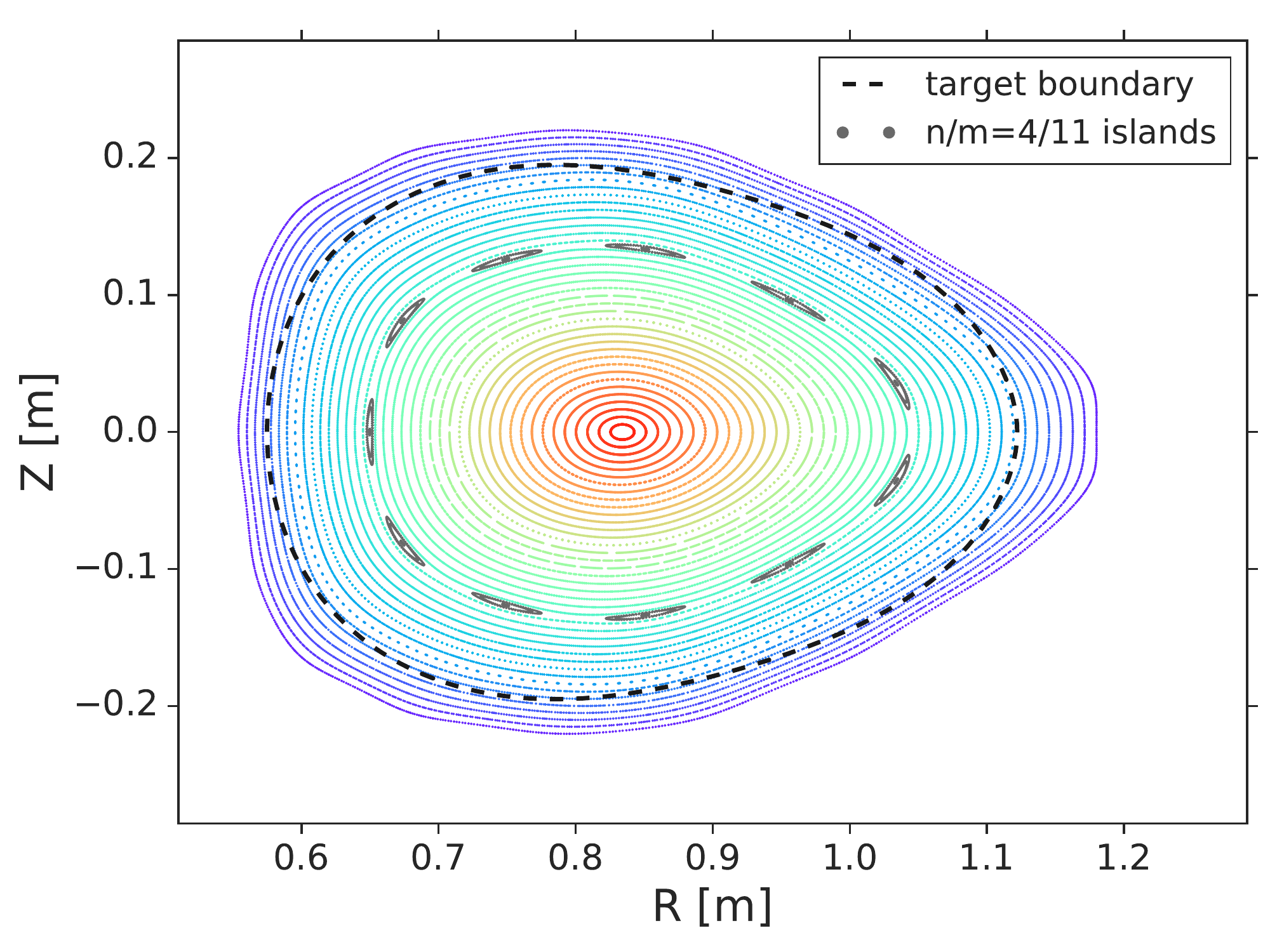}
    \caption{Shape of the target plasma boundary (black-dashed) and \Poincare plots (colorful dots) of field line tracing at the bullet-shaped cross-section. Field lines starting at $r \in [0.83, 1.20], z=0.0$ are followed 1000 periods. The $n/m=4/11$ islands are highlighted with grey dots.}
    \label{fig:IdealPP}
\end{figure}

Although multiple scenarios have been investigated with different beta values \cite{SHIMIZU2018}, CFQS and its modular coil system \change{was}{were} optimized by targeting a zero-beta (vacuum) configuration.
For simplicity, we only consider the vacuum configuration for the following calculations.
The target plasma boundary at the bullet-shaped cross-section and \Poincare plots by tracing field lines in the magnetic field produced by the designed coils are shown in \Fig{IdealPP}.
It should be noted that the magnetic field is calculated from single filaments located at the center of each coil. 
In this paper, we will ignore the effect of finite-build coils and only use coil filaments for the following calculation (although coils are plotted with finite size).
Note that the target boundary is not exactly matched in \Fig{IdealPP}.
There are two reasons.
First, the desired magnetic field is not the only target when designing CFQS coils. 
The average residual $\vect{B} \cdot \vect{n} / |\vect{B}|$ on the target boundary is about 1\%. 
Secondly, \change{}{the} coils used in this paper are slightly different from the original ones, as FOCUS does not directly use the raw points in space.
FOCUS adopts Fourier representation to describe coils and truncation errors exist when fitting the actual coil data.
Hereafter, CFQS (modular) coils are referred to single filaments described with Fourier representation in FOCUS.

\subsection{n/m=4/11 magnetic islands} \label{result_rp}
\begin{figure}
    \centering
    \includegraphics[width=0.8\textwidth]{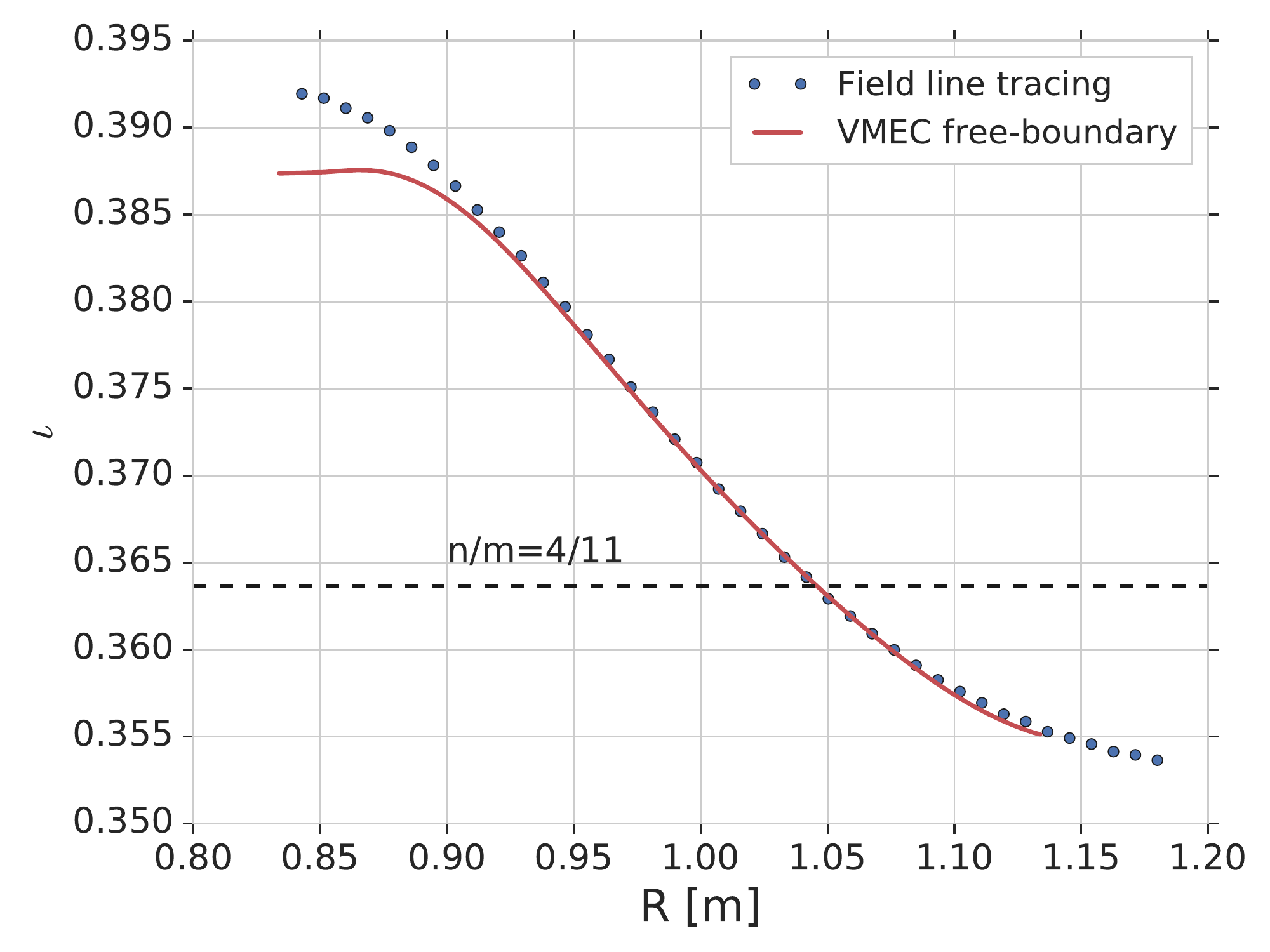}
    \caption{Rotational transform ($\iota$) profiles from VMEC free-boundary run (red solid) and field line tracing (blue dots). The abscissa is the radial position of flux surfaces at the bullet-shaped cross-section as in \Fig{IdealPP}. The discrepancy near the axis might be caused by bad convergence on those flux surfaces in VMEC free-boundary run.}
    \label{fig:iota}
\end{figure}

CFQS is a low shear configuration.
The rotational transform profile was carefully chosen to avoid low-poloidal-number islands.
Rotational transform profiles from VMEC free-boundary calculation with the designed coils and from field line tracing are shown in \Fig{iota}.
The $n/m=4/11$ island chain is found to have the lowest poloidal number inside the boundary.
From \Fig{IdealPP}, one can clearly observe the 4/11 islands, although their sizes are small.
We shall select $b_{11,4}$ as our reference resonant perturbation and the present value of $b_{11,4}$ is served as the target value $b^o_{11,4}$.
The metric described in \Eqn{frp} is now evaluating how far the 4/11 islands are away from the present size. 
With the designed coils, $F_{RP} = 0$ is at a global minimum.
The 4/11 island chain might not be serious (because of relatively large $m$), but it would be sufficient for demonstrating the new Hessian matrix method.

To approximate the $\iota=4/11$ rational surface, free-boundary VMEC was executed with relatively high radial resolution (256 surfaces with maximum Fourier modes Mpol=10, Ntor=20).
Afterwards, the flux surface that is closest to $\iota=4/11$ surface was selected and the code BOOZ\_XFORM \cite{Sanchez2000} was used to convert VMEC coordinates into Boozer coordinates.
FOCUS then read the reference flux surface \change{represented in Boozer coordinates}{parameterized in Boozer angles ($\t_B, \z_B$)}.
\change{}{The magnetic field $\vect{B}_0$ was calculated from the actual coils, as the reference configuration is zero-beta.
The resonant harmonic $b_{11,4}$ and its derivatives (both first- and second-order ones) were then calculated by following \Eqn{bmn}, \Eqn{frp_deriv1} and  \Eqn{frp_deriv2}, where we fixed the flux surface and the Boozer coordinates on it.}
\change{Details of how FOCUS handles Boozer coordinates are shown in appendix \ref{booz_xform}.}{}

When using Fourier representation in FOCUS, the vector of coil parameters consists of Fourier coefficients and coil currents,
\be \label{eq:fourier}
\vect{X} \equiv \{ x^c_{k,n}, x^s_{k,n}, y^c_{k,n}, y^s_{k,n}, z^c_{k,n}, z^s_{k,n}, I_k \} \ ,
\ee
where $x^c_{k,n}$ ($x^s_{k,n}$) is the $n$-th cosine (sine) Fourier coefficient of
the $k$-th coil and $I_k$ is the current.
To accurately describe CFQS modular coils, we have performed a scan of the number of Fourier coefficients used in each spatial coordinate.
According to appendix \ref{nf_scan}, the optimal value of the maximum Fourier mode $N_F$ is 10.
The total number of degrees of freedom is 1024 since each modular coil is considered to be independent (as they will be fabricated and assembled separately). 
Coil currents are normalized to the average current, $3.125 \tento{5}$ A, while Fourier coefficients are normalized to a geometric quantity, $1.23$ m.

\begin{figure}[ht]
    \centering
    \includegraphics[width=0.7\textwidth]{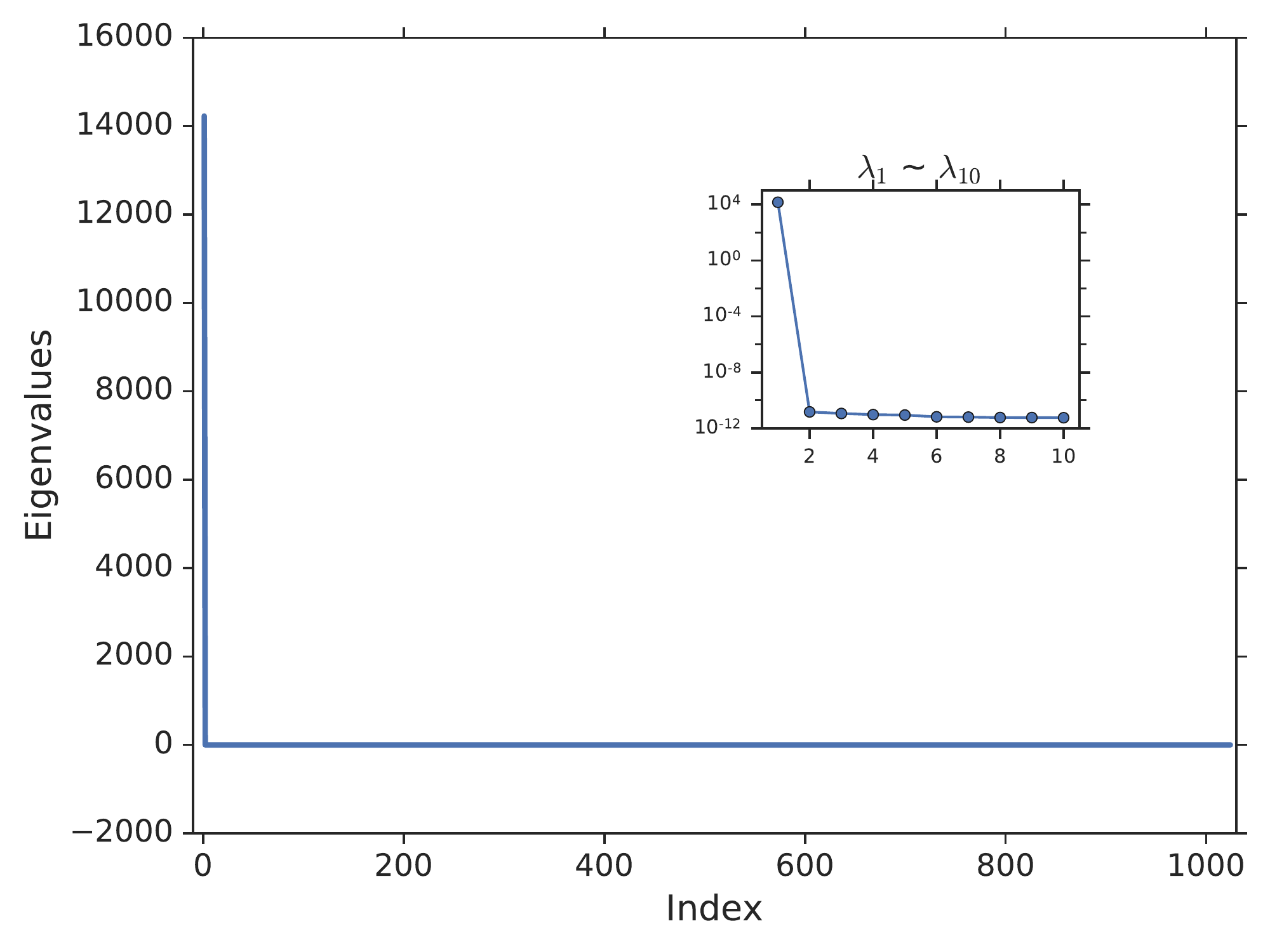}
    \caption{Eigenvalues of the Hessian matrix for the size of 4/11 islands. All eigenvalues are sorted in descending order with $\lambda_1 = 1.42 \tento{4}$ and $\lambda_{1024} = -1.80 \tento{-10}$.}
    \label{fig:eigenvalue_rp}
\end{figure}

\begin{figure}[ht]
    \centering
    \includegraphics[width=0.8\textwidth]{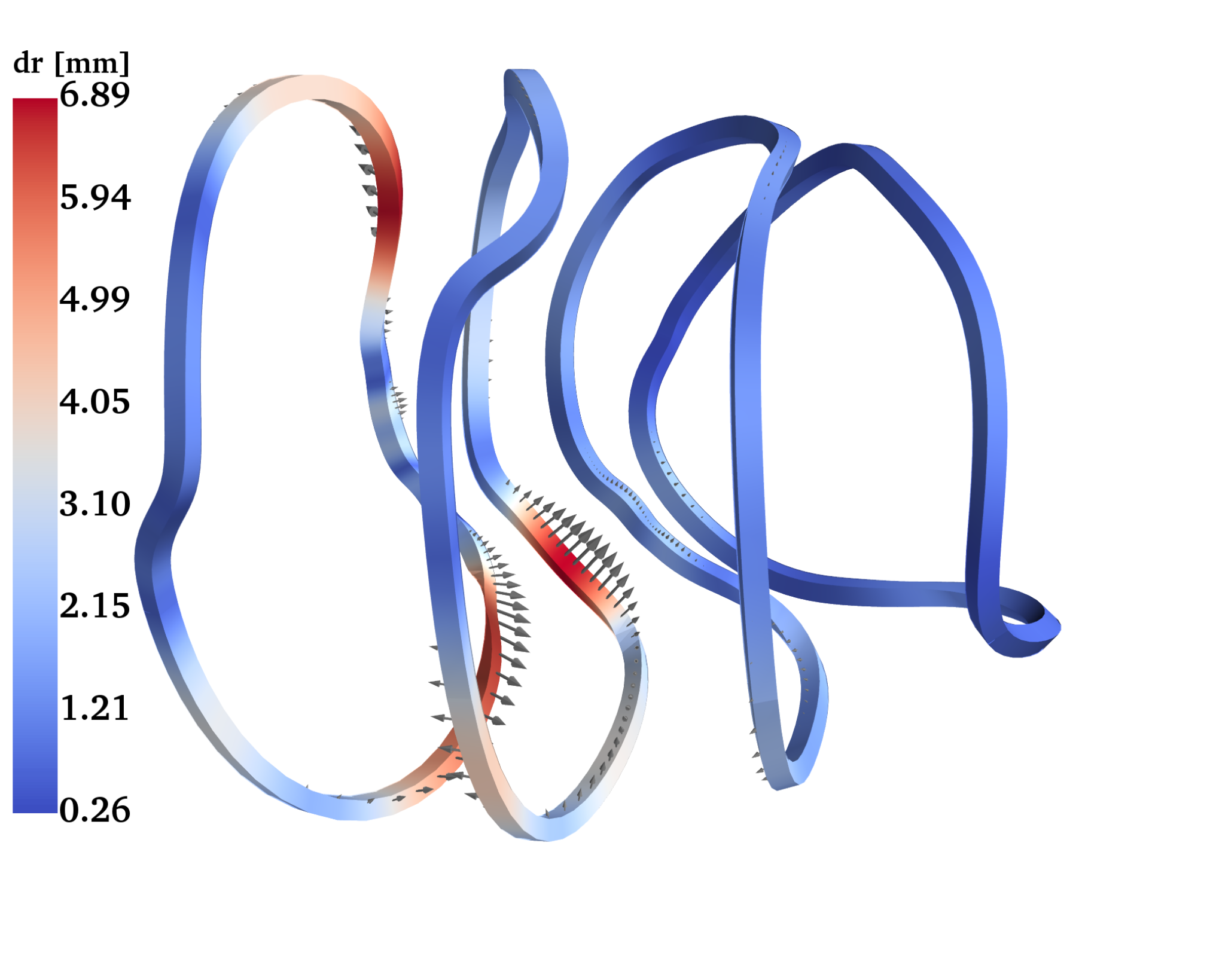}
    \caption{Coil deformations under the first principal eigenvector for 4/11 islands ($\vect{X}_0 + \xi \vect{v}_1$). Perturbation magnitude $\xi$ is 0.01. Colors on coils imply the departure distance of data points in each coil and arrows demonstrate the deforming direction.}
    \label{fig:ev1_rp}
\end{figure}

The Hessian matrix of the 4/11 islands is then computed by FOCUS. (The computation time is about 100 seconds with 64 cores.)
Eigenvalue decomposition is carried out and \Fig{eigenvalue_rp} shows the spectrum of eigenvalues in descending order.
Only the first (largest) eigenvalue has a significant value and all the others are almost zero.
This implies that the Hessian matrix is positive semi-definite, which is no surprise as $F_{RP}$ is at an exact global minimum.
The first principal eigenvector, which is associated with the largest eigenvalue, will have the highest impact on the size of 4/11 islands. 
It represents the most \change{dangerous}{important} perturbation scheme in all the possible combinations with the same magnitude in parameter space.
\Fig{ev1_rp} illustrates how the coils will deform if perturbed in the direction of the first principal eigenvector with a finite magnitude ($\xi=0.01$).
It is interesting that the first principal eigenvector is preserving periodicity and stellarator symmetry, although all the 16 modular coils are considered independently.
Coil currents are also free variables and the currents will be perturbed by amounts of $\xi \times (-2160.67, -1949.13, 1877.62, 2529.56)$ A in each type of coil, which are negligible compared to the actual coil current ($3.125\tento 5$A).
Results in \Fig{ev1_rp} suggest that the M1 \& M2 coils are much more sensitive than M3 \& M4. 
Particularly, toroidal deformations at three regions, inboard upper corner, inboard lower corner of M1 and inboard part under the middle plane of M2, have the most significant effects on preserving 4/11 islands.

\begin{figure}
    \centering
    \begin{subfigure}[ht]{0.48\textwidth}
    \centering
    \includegraphics[width=\textwidth]{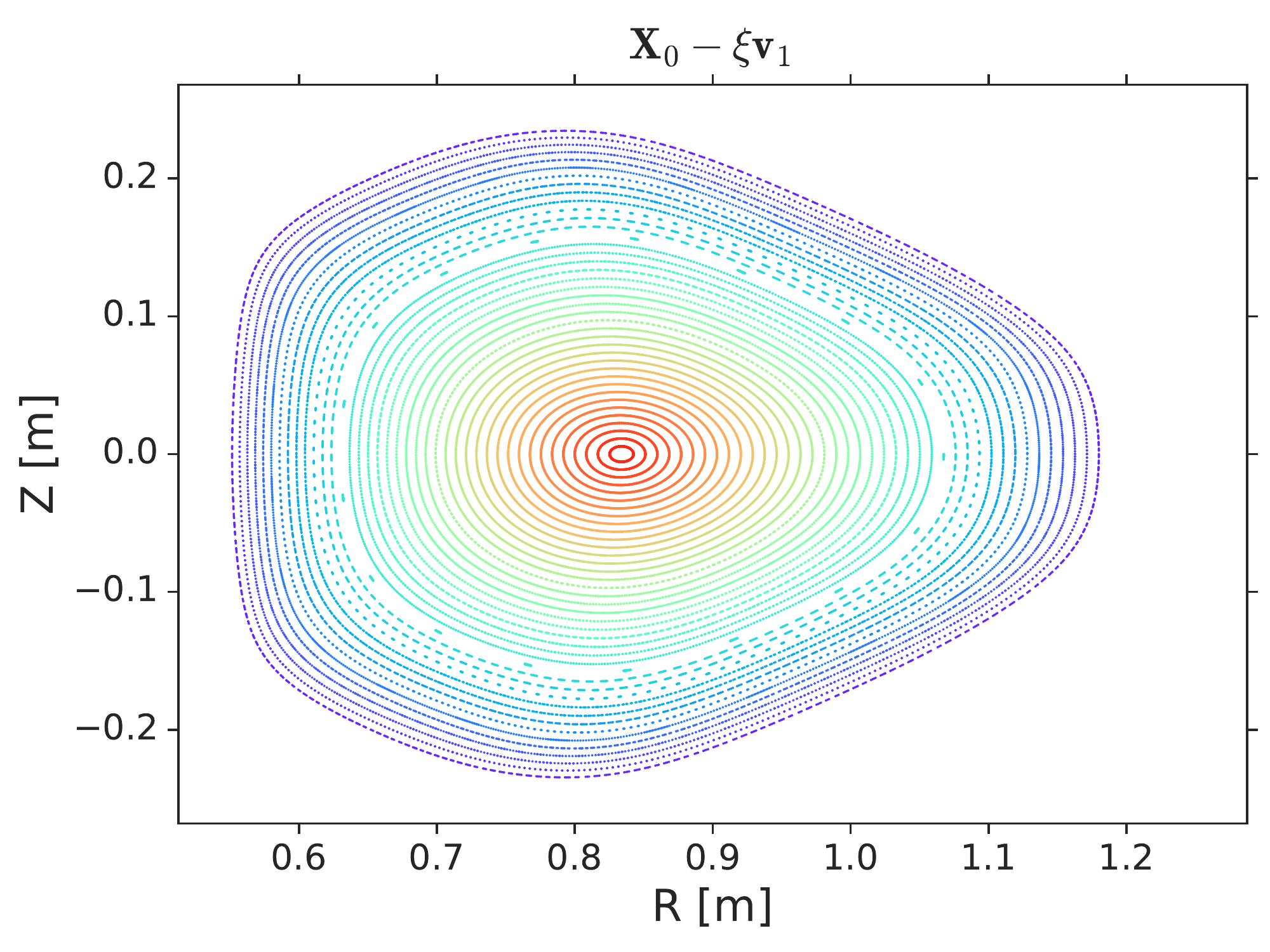}
    \caption{Negatively perturbed coils, islands eliminated;}
    \label{fig:ev1n_rp_pp}
    \end{subfigure} %
    \begin{subfigure}[ht]{0.48\textwidth}
    \centering
    \includegraphics[width=\textwidth]{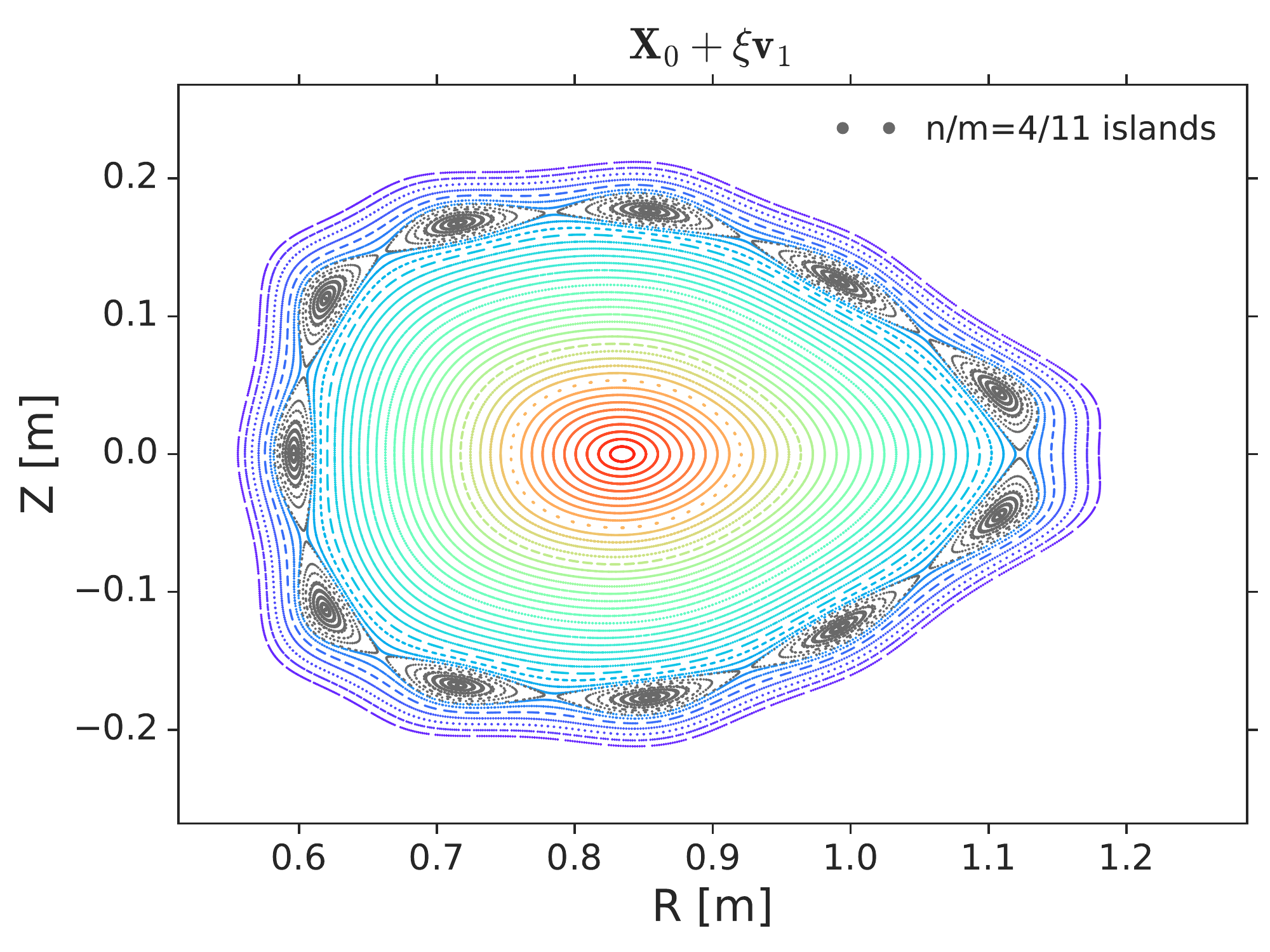}
    \caption{Positively perturbed coils, islands enlarged.}
    \label{fig:ev1p_rp_pp}
    \end{subfigure}
    \caption{\Poincare plots of perturbed coils in two different directions following the first principal eigenvector (for the size of 4/11 islands) with the same magnitude $\xi=0.01$. The 4/11 islands are either eliminated in (a) or enlarged in (b), compared to \Fig{IdealPP}. Numerical details for field line tracings are the same as \Fig{IdealPP}.}
    \label{fig:ev1_rp_pp}
\end{figure}

One can verify the effects of coil deviations by checking the size of 4/11 islands.
Field line tracings are performed with the perturbed coils.
\Poincare plot of the negatively perturbed coils ($\vect{X}_0 - \xi \vect{v}_1$) is shown in \Fig{ev1n_rp_pp} and the positively  perturbed one ($\vect{X}_0 + \xi \vect{v}_1$) is in \Fig{ev1p_rp_pp}, while $\xi=0.01$ in both cases.
As shown in \Fig{ev1_rp_pp}, the 4/11 islands are either suppressed or enlarged when perturbing the coils in the two directions.
Both eliminations and enlargements of the islands are deteriorating our figure of merit since we provide a non-zero $b^o_{11,4}$ in \Eqn{frp} and it will preserve a finite island chain.

\subsection{Quasi-axisymmetry} \label{result_qa}
\begin{figure}[ht]
    \centering
    \includegraphics[width=0.8\textwidth]{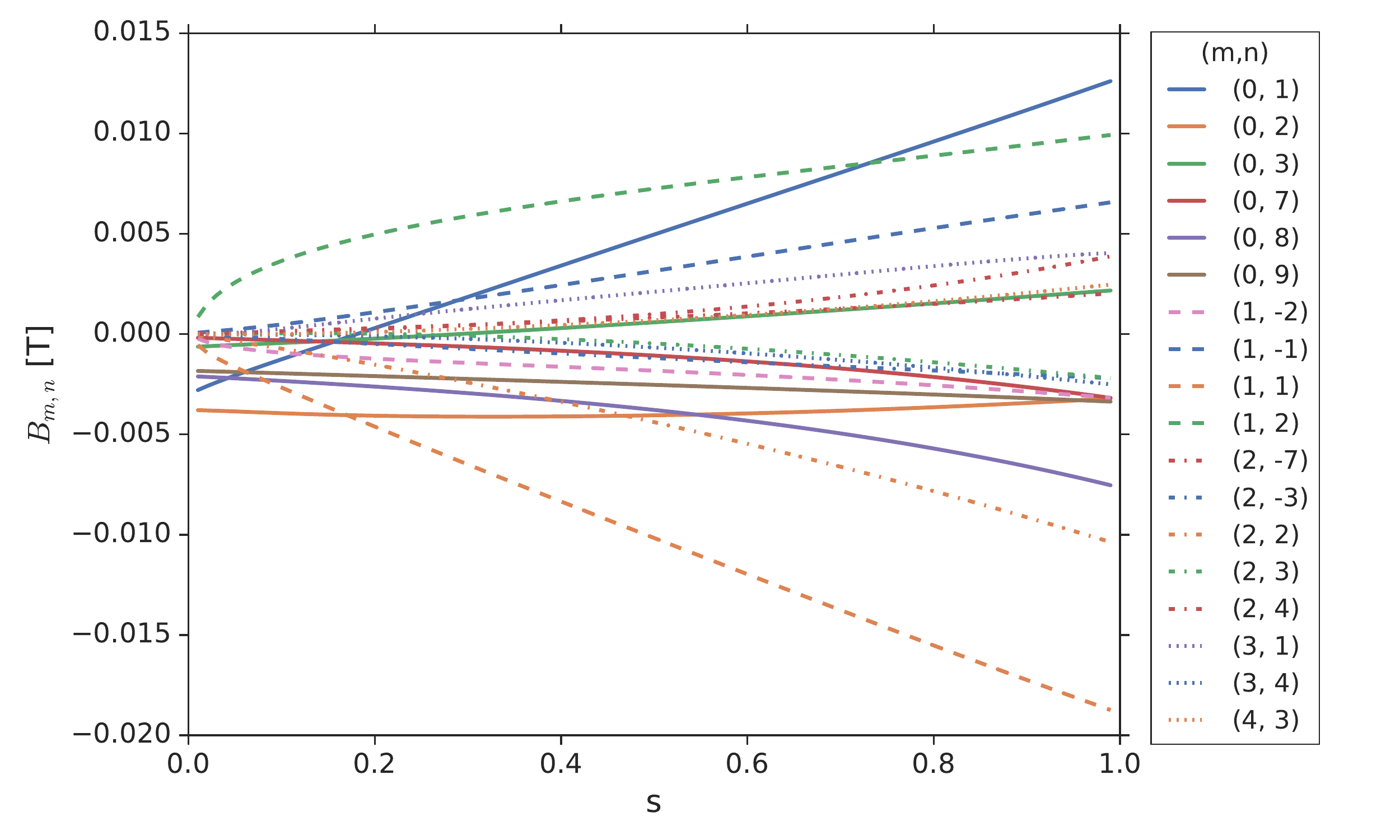}
    \caption{Spectra of Fourier components $B_{m,n}$ for the magnetic field strength in CFQS configuration. Fourier components are calculated by BOOZ\_XFORM (Mpol=100, Ntor=30) with free-boundary VMEC equilibrium using the designed coils. The abscissa is the normalized toroidal flux, $s=\Psi / \Psi_{edge}$. Only the terms that are larger than 0.002 T at the edge are shown.}
    \label{fig:qa_actual}
\end{figure}

Similarly, quasi-axisymmetry of the magnetic field could be our figure of merit.
By assessing the non-axisymmetric terms ($n \neq 0$) in \Eqn{fqs}, we could evaluate the QA quality of the produced magnetic field.
\Fig{qa_actual} shows the spectra of main non-zero non-axisymmetric terms from BOOZ\_XFORM converting free-boundary VMEC results with the original CFQS coils.
Because the flux surface is stellarator symmetric, odd (sin) components in $B_{m,n}$ are zero and hereafter we only consider even (cosine) components.
The leading non-axisymmetric terms, $B_{0,1}$, $B_{1,1}$, $B_{1,2}$ and $B_{2,2}$, are significantly smaller than $B_{0,0}$ ($\sim 0.95$T).
QA quality, as evaluated in \Eqn{fqs}, monotonically degrades from the axis to edge.
We can select an arbitrary flux surface, \eg the flux surface at $s=\Psi / \Psi_{edge} = 0.5$, as the reference surface.
After reading in flux surface information and Boozer coordinates, FOCUS calculates the magnetic field produced by coils on the reference surface and then Fourier coefficients $B_{m,n}$ are gauged by using \Eqn{Bmn}.
The maximum poloidal and toroidal mode numbers used here are Mpol=4, Ntor=8 ( $m \in [0, \mathrm{Mpol}]$ and $n \in [-\mathrm{Ntor}, \mathrm{Ntor}]$).

\begin{figure}[ht]
    \centering
    \includegraphics[width=0.7\textwidth]{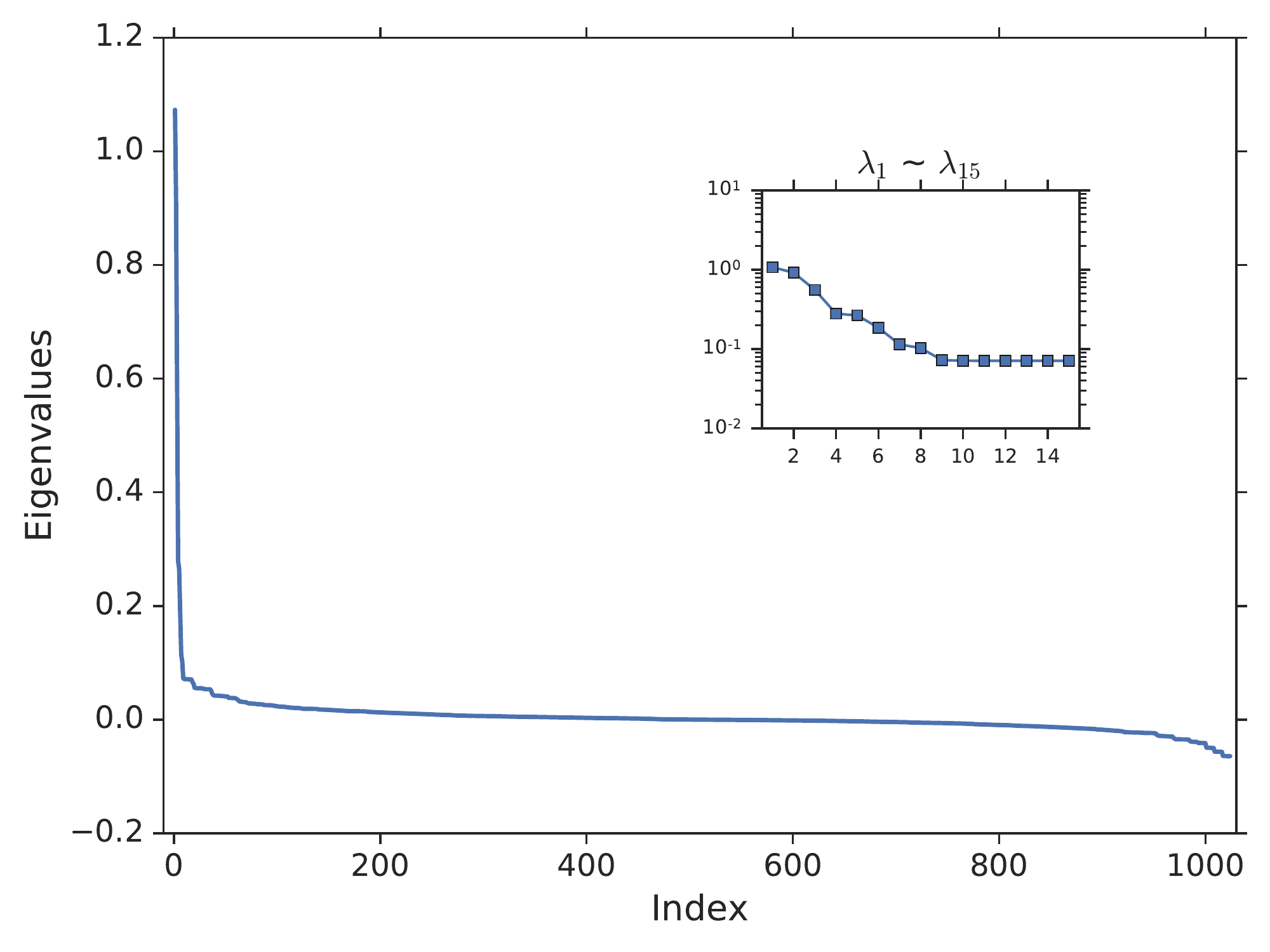}
    \caption{Eigenvalues of the Hessian matrix for QA quality. All eigenvalues are sorted in descending order with $\lambda_1 = 1.07 $ and $\lambda_{1024} = 6.42\tento{-2}$. The last 515 eigenvalues are negative. }
    \label{fig:eigenvalue_qa}
\end{figure}

Eigenvalues of the Hessian matrix regarding \change{to}{} QA quality are shown in \Fig{eigenvalue_qa}.
Numerical details are the same as in Sec. \ref{result_rp}.
Unlike the 4/11 island size, the eigenvalues for QA quality on the half-toroidal-flux surface are more divergent.
The four largest eigenvalues ($\lambda_1$, $\lambda_2$, $\lambda_3$, $\lambda_4$) are of values ($1.07$, $0.92$, $0.56$, $0.28$), while the smallest one $\lambda_{1024} = -6.42\tento{-2}$.
Actually, there are 515 negative eigenvalues.
This implies that the Hessian matrix is not positive \change{}{semi-}definite and the figure of merit (QA quality) used here is not at a \change{strict}{local} minimum. 
It is expected since the coils are not designed by targeting quasi-axisymmetry (on this particular flux surface) and there is a trade-off for finding easy-to-build coils.
However, the quadratic approximation in \Eqn{quadratic} is still valid, as checked in appendix \ref{quadratic}.
QA quality is marginally close to a minimum.
Therefore, we can keep using the Hessian matrix method.
\change{}{Here, the target value for QA quality was set to be zero, \ie CFQS would like to have as high QA quality as possible.
This is not essential and we could modify it to include a non-zero target value like in \Eqn{frp}.}

\begin{figure}
    \centering
    \begin{subfigure}[ht]{0.48\textwidth}
    \centering
    \includegraphics[width=\textwidth]{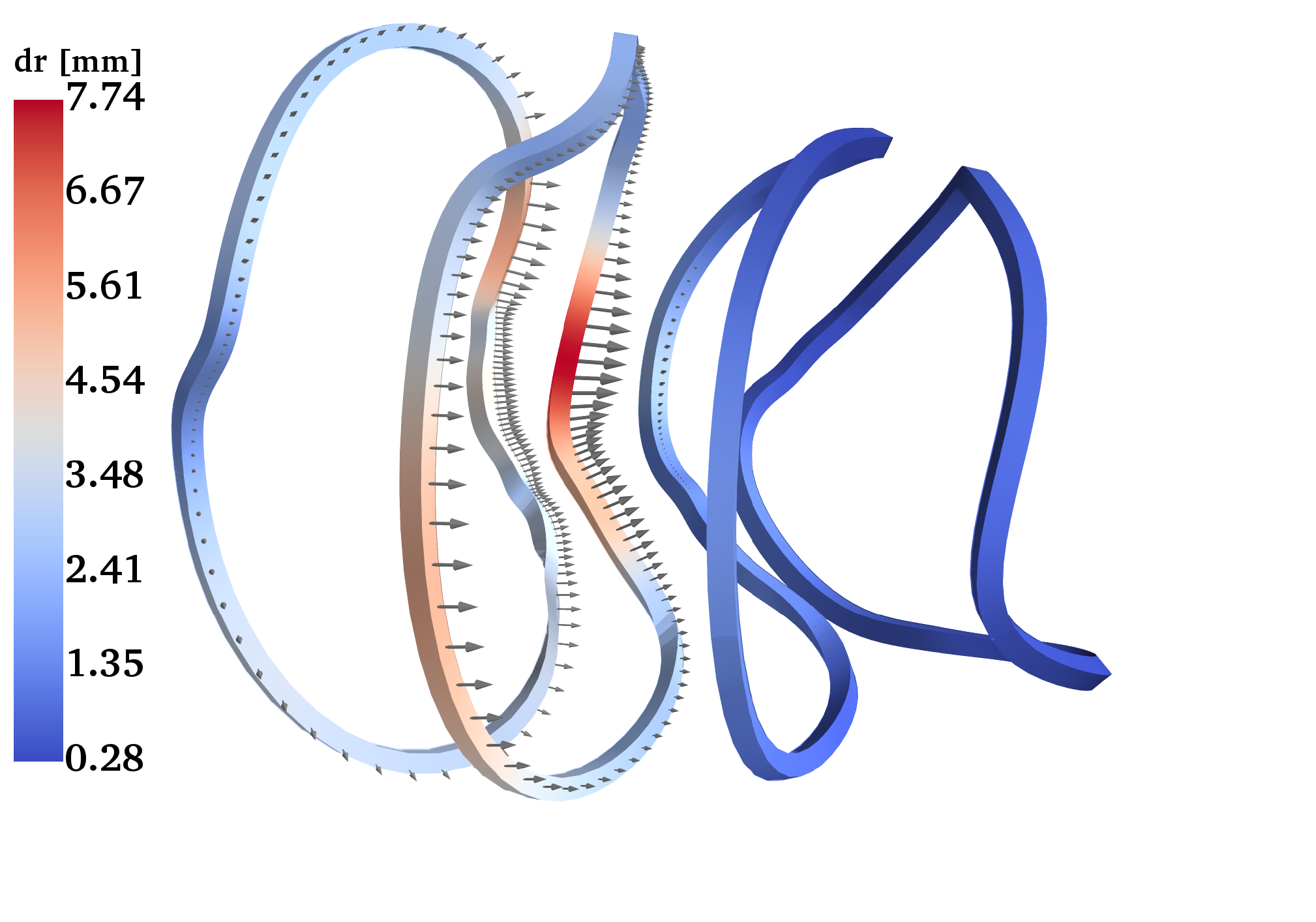}
    \caption{$\vect{X}_0 + \xi \vect{v}_1$ ($\lambda_1 = 1.09$)}
    \label{fig:ev1_qa}
    \end{subfigure} %
    \begin{subfigure}[ht]{0.48\textwidth}
    \centering
    \includegraphics[width=\textwidth]{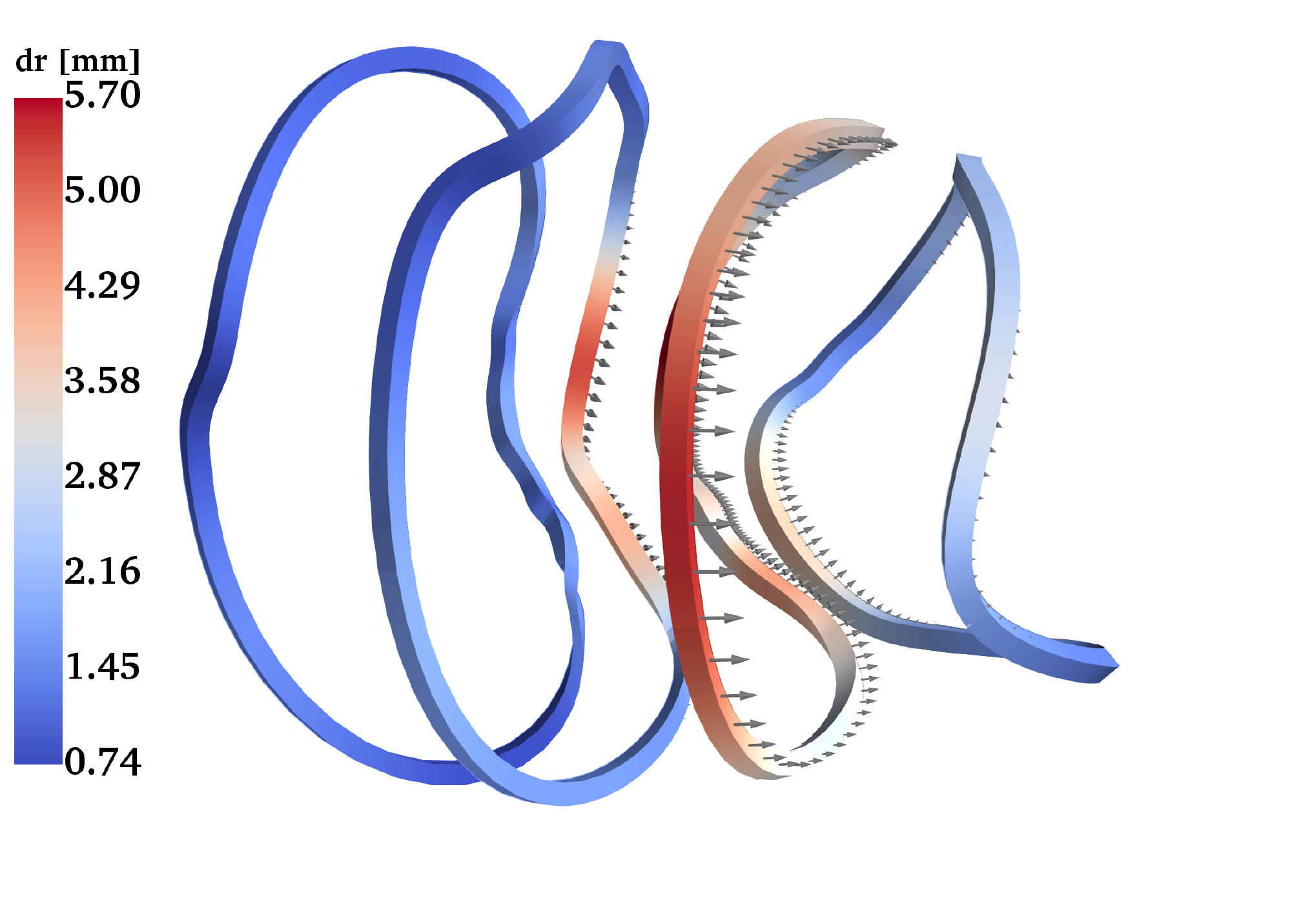}
    \caption{$\vect{X}_0 + \xi \vect{v}_2$ ($\lambda_2 = 0.92$)}
    \label{fig:ev2_qa}
    \end{subfigure}
    \bigskip  
    \begin{subfigure}[ht]{0.48\textwidth}
    \centering
    \includegraphics[width=\textwidth]{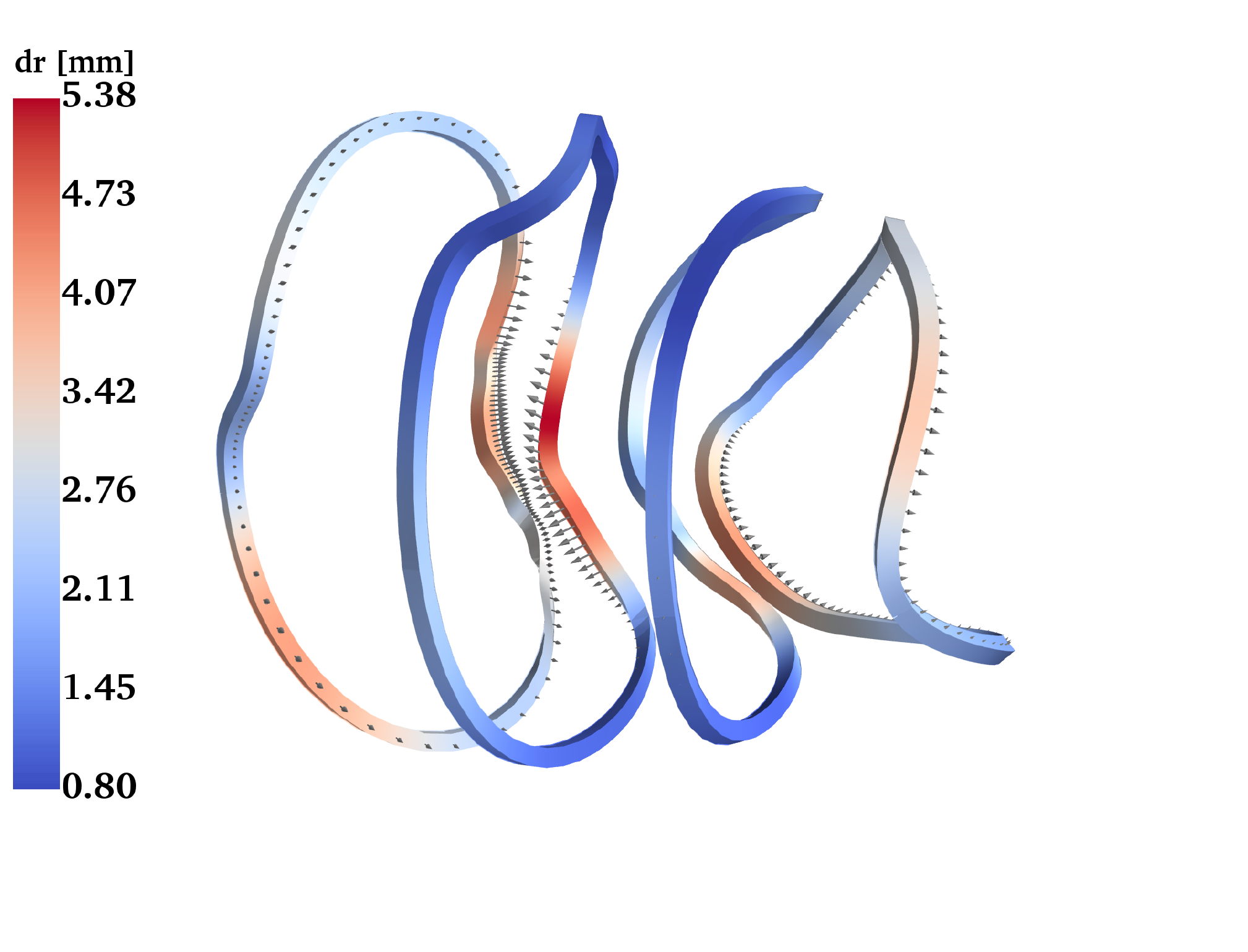}
    \caption{$\vect{X}_0 + \xi \vect{v}_3$ ($\lambda_3 = 0.56$)}
    \label{fig:ev3_qa}
    \end{subfigure} %
    \begin{subfigure}[ht]{0.48\textwidth}
    \centering
    \includegraphics[width=\textwidth]{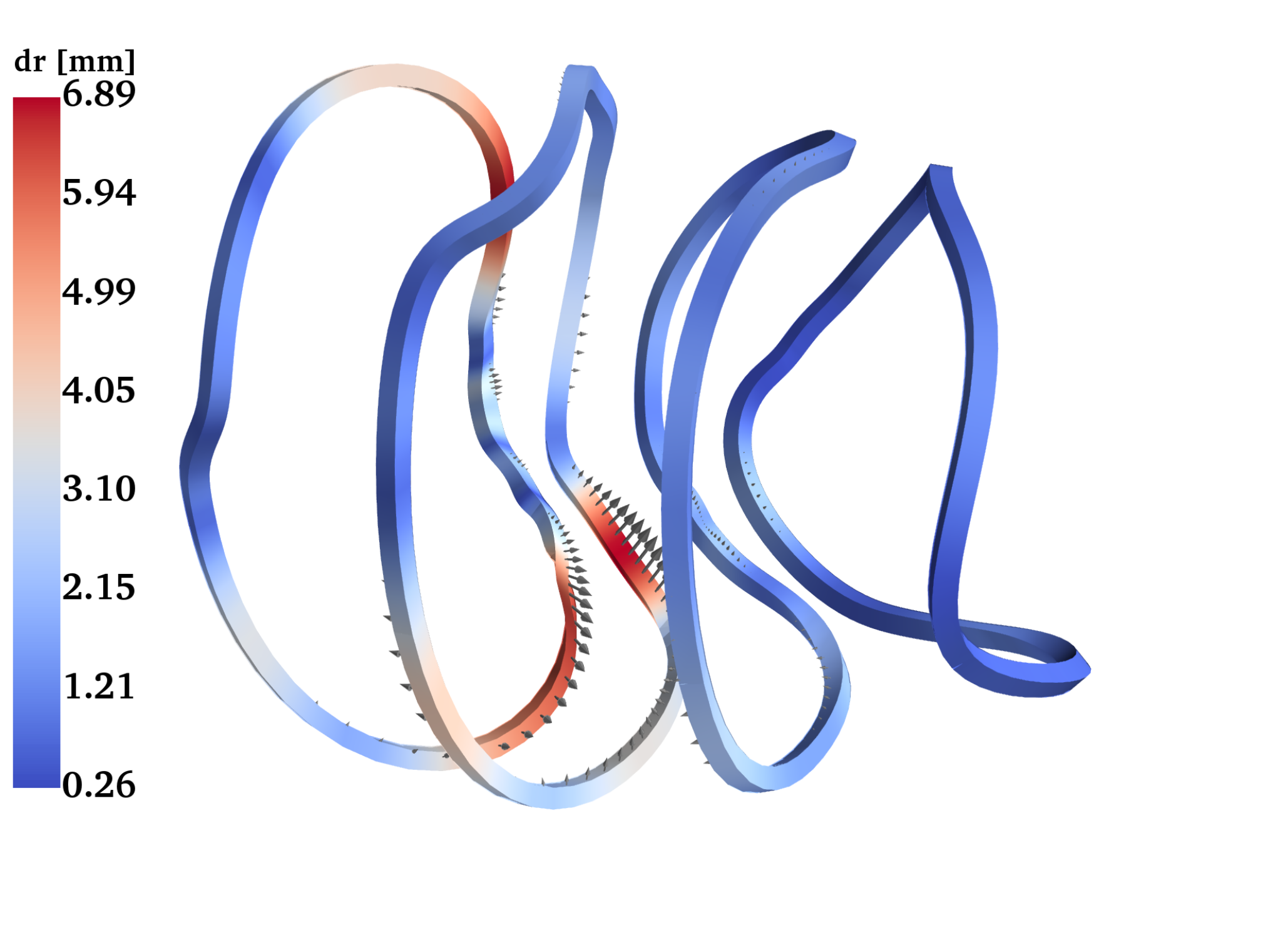}
    \caption{$\vect{X}_0 + \xi \vect{v}_4$ ($\lambda_4 = 0.28$)}
    \label{fig:ev4_qa}
    \end{subfigure}   
    \caption{Coil deformations under the first four principal eigenvectors for QA quality on the half-toroidal-flux surface. Perturbation magnitudes $\xi$ are all set to 0.01. Colors on coils imply the departure distance of data points in each coil and arrows demonstrate the deforming direction.}
    \label{fig:ev_qa_coils}
\end{figure}

Likewise, the effects of eigenvectors on coils can be studied by mapping the perturbed coils back to real space.
This time, we need to consider more than one eigenvector.
In \Fig{ev_qa_coils}, how coils deform under the first four principal eigenvectors with the same size perturbation $\xi=0.01$ (in positive directions) are displayed.
Currents flowing in the coils are also varied but \change{limited}{within} a small range (less than 0.09\% of the original current).
Again, perturbed coils are still following stellarator symmetry and periodicity, so only one fourth of the coils are shown.
The maximum departures in the four schemes are not identical, but close.
Distributions of highly deformed regions are also different.

\section{Coil tolerance calculation} \label{result_tol}
Once the eigenvalues and eigenvectors of the Hessian matrix have been obtained, coil tolerance can be computed.
The change in the target figure of merit is quadratically approximated, as shown in \Eqn{quadratic}.
\change{}{Suppose we have an arbitrary perturbation $\ds \Delta \vect{X} = \sum_i^N a_i \vect{v}_i$ where $a_i$ is the $i$-th component in the direction of $\vect{v}_i$.
The resulting error field can be approximated by $\ds \Delta F \approx \sum_i^N \half \lambda_i {a_i}^2$.
Since the eigenvalues are in descending order ($\lambda_1 \geq \lambda_2 \geq \cdots \geq \lambda_N$), \changenew{for a given magnitude of perturbation, $\ds \abs{\Delta \vect{X}}=\sqrt{\sum_i^N {a_i}^2}=\xi$, the largest error field happens in the direction of the first principal eigenvector ($\Delta \vect{X} = \xi \vect{v}_1$, $\Delta F=\half \lambda_1 \xi^2$).}{for all the perturbations that have the same magnitude ($\ds \abs{\Delta \vect{X}}=\sqrt{\sum_i^N {a_i}^2}=\xi$), the one that only consists of the first principal eigenvector ($\Delta \vect{X} = \xi \vect{v}_1$) will generate the largest error field ($\Delta F=\half \lambda_1 \xi^2$).}
}
If we have a maximum allowable deterioration on the figure of merit, \ie ${\Delta F}_{tol}$, the associated tolerance of the perturbation magnitude in the direction of \change{an arbitrary}{the first principal} eigenvector is calculated as,
\be \label{eq:tol}
\xi_{tol} \approx \sqrt{2 \ {\Delta F}_{tol} / \lambda_1} \ .
\ee
\change{The first principal eigenvector, which represents the most dangerous coil deformation, would have the tightest tolerance in parameter space.
After mapping to real space, we can obtain the worst coil deviation scheme and the allowable perturbation magnitude.}
{Here, the perturbation magnitude is calculated in parameter space (\eg Fourier space).
More useful information for engineers would be the actual coil tolerance in real space (values in $mm$). 
We can obtain the worst coil deviation scheme and the allowable perturbation magnitude in real space by mapping the perturbed coils back to the real space.
Discrepancies on values after the transformation are expected, especially when we use Fourier representation for coils.
Other local representations, like piece-wise linear representation and splines, might be able to resolve this discrepancy. 
In this paper, we are going to stay with Fourier representation and conduct some initial attempts to calculate coil tolerance in real space.}

For CFQS, as stated above, the rotational transform profile is carefully chosen to avoid low-rational-number islands.
Therefore, resonant magnetic islands are not the first priority for the error field study.
Meanwhile, the primary goal of the device is to demonstrate the properties of quasi-axisymmetry.
We shall only use the QA quality on the half-toroidal-flux surface to calculate coil tolerances.

If we assume the maximum acceptable QA deterioration is 10\% of the present value ($F_{QA} = 2.95\tento{-4}$), \change{the allowable perturbation magnitudes calculated using \Eqn{tol} in the first four principal eigenvectors are $7.42\tento{-3}$, $8.02\tento{-3}$, $1.03\tento{-2}$ and $1.45\tento{-2}$.}{the magnitude of the allowable perturbation calculated using \Eqn{tol} in the first principal eigenvector is  $7.42\tento{-3}$.}
Note that for \change{the same magnitude of perturbations}{perturbations having the same magnitudes in Fourier space} \change{(in parameter space)}{}, coil deviations in real space are not of the same amount (albeit they are of the same order), as shown in \Fig{ev_qa_coils}.
\change{In this case, since the differences of the four largest eigenvalues are not substantial, the discrepancy of perturbation magnitude in parameter space and in real space for each eigenvector cannot be negligible.}
{In this case, the magnitudes of perturbation in the other three principal eigenvectors are $8.02\tento{-3}$, $1.03\tento{-2}$ and $1.45\tento{-2}$.
Since the differences of the four largest eigenvalues are not substantial, we will try to compare all the four associated eigenvectors.} 
\change{If we consider coil tolerance in real space,}{If we only consider perturbations in the direction of eigenvectors,} the allowable coil deviations for the first four eigenvectors are 5.74, 4.57, 5.54 and 10.02 mm.
Detailed coil deformations can be visualized in \Fig{ev_qa_coils} (using allowable perturbation magnitudes).
A maximum allowable deviation of about 5 mm is much more generous than coil tolerance in NCSX and W7-X, especially considering that CFQS has a smaller size.

Here, the value of allowable 10\% of QA quality deterioration is \change{}{chosen }just for demonstration.
Serious numbers can be assessed by comprehensive calculations, like evaluating the $\epsilon_{eff}$ or do particle transport simulations, and it is beyond the scope of this paper.
\change{}{In the above calculations, we were evaluating the first four principal eigenvectors and the contributions from coil currents were all small.
For obtaining coil tolerance in real space, coil currents should probably be excluded from free coil parameters.}

\section{Improve coil designs towards better physics properties} \label{QA_opt}
The two figures of merit mentioned in this paper, magnetic islands and quasi-symmetry, are two important physics properties.
In addition to quantifying error fields, they can be used to optimize coils for better physics properties.
For instance, the 4/11 islands in CFQS can be eliminated if the coil shapes are varied following a certain direction, as shown in \Fig{ev1n_rp_pp}.
More generally, specific islands could be reduced if we minimize \Eqn{frp} by targeting zero island width.
This idea was successfully applied to NCSX by Hudson \etal \cite{Hudson2002}.

\begin{figure}[ht]
    \centering
    \includegraphics[width=0.8\textwidth]{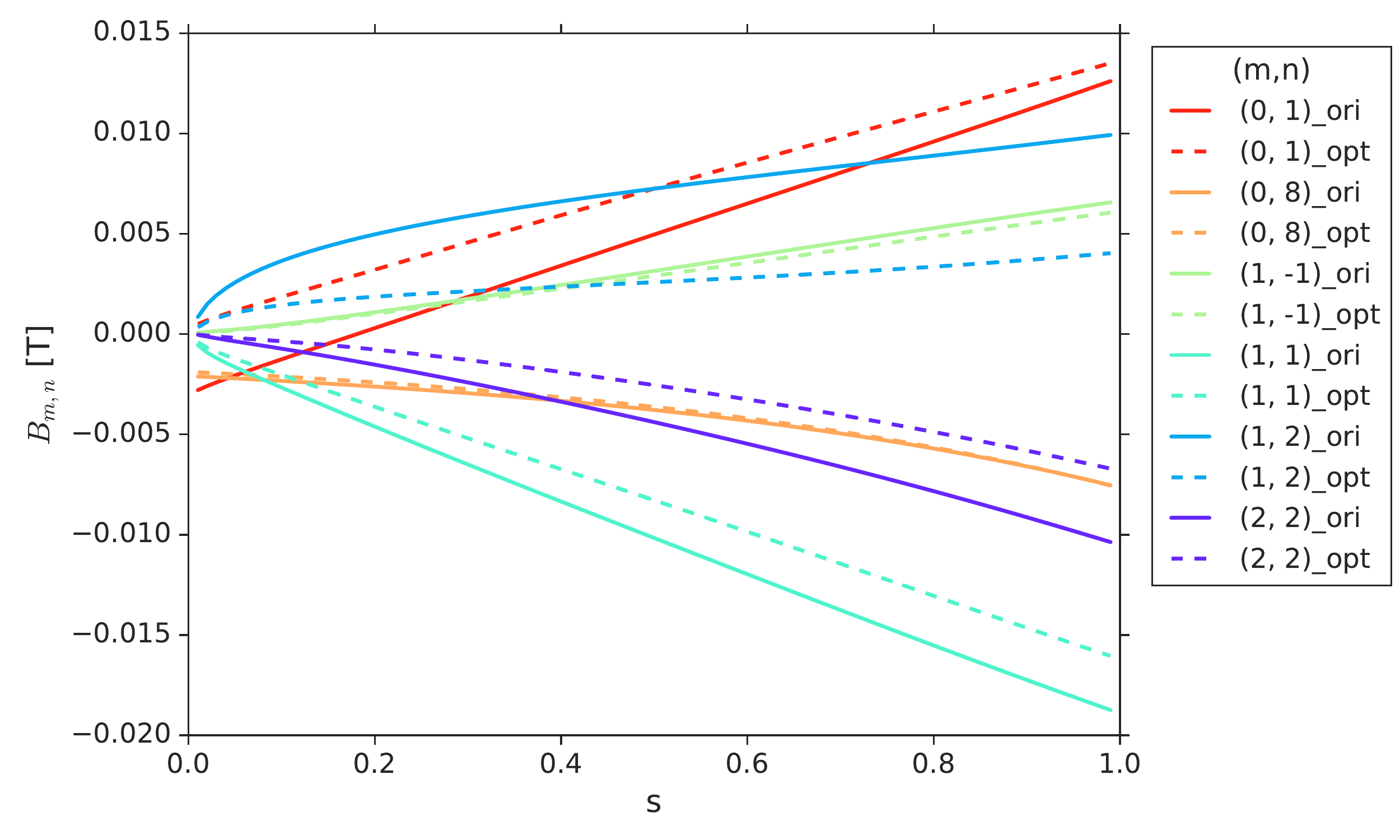}
    \caption{Comparison of leading non-axisymmetric terms between original (solid) and optimized (dashed) CFQS coils. $B_{m,n}$ is calculated by BOOZ\_XFORM from free-boundary VMEC runs. Only the terms that are larger than 0.005 T at the edge are shown.}
    \label{fig:qa_compare}
\end{figure}

\begin{figure}[ht]
    \centering
    \includegraphics[width=0.75\textwidth]{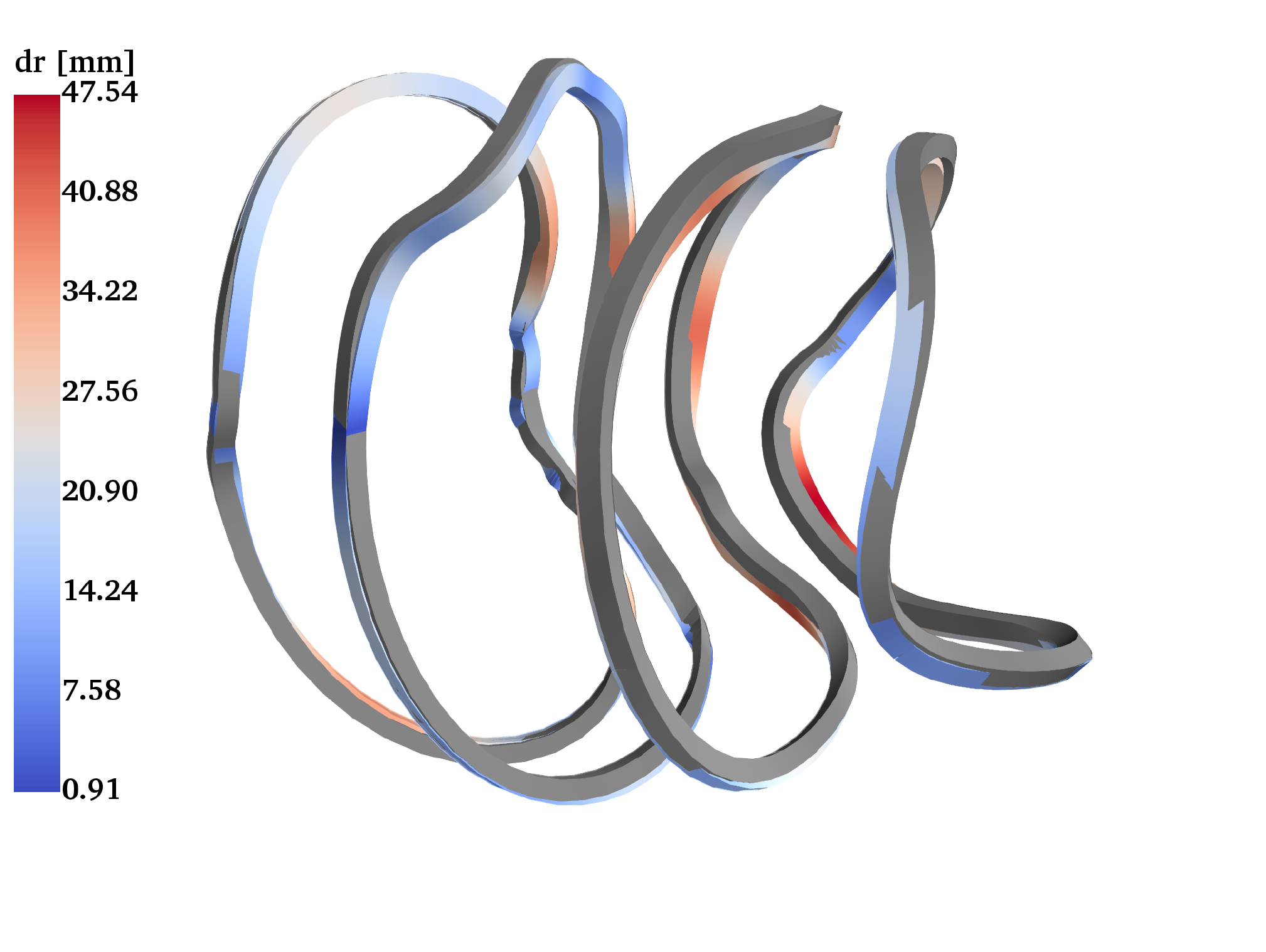}
    \caption{QA optimized coils (colored) and the original CFQS coils (grey). Only one fourth of the sixteen coils are shown. Colors imply the distance of data points between the original and optimized coils. The coils are plotted with rectangle cross-sections ($50$mm $\times 50$mm).}
    \label{fig:QA_opt_coils}
\end{figure}

Here, we will show that we can also improve the quasi-axisymmetry quality of CFQS.
As shown in \Fig{eigenvalue_qa}, there are several negative eigenvalues of the Hessian matrix.
If the coils are varied following the eigenvectors associated with negative eigenvalues, the change of the functional in \Eqn{fqs} will be negative as well.
This indicates that the QA quality on the half-toroidal-flux surface will be improved.
Instead of varying coils manually, we can directly optimize the QA quality as one of the cost functions in FOCUS.
A simple optimization was carried out by targeting only the cost function of QA quality on the half-toroidal-flux surface.
The actual CFQS coils were used as the initial guess.
After 10 iterations using a conjugate gradient method, a new coil set was found.
\change{}{$F_{QA}$ was reduced from $2.95\tento{-4}$ to $4.42\tento{-5}$ on the half-toroidal-flux surface.}
Free_boundary VMEC with the new coils and BOOZ\_XFORM calculations were then conducted.
In \Fig{qa_compare}, we show the leading non-axisymmetric terms from the optimized coils compared with the original coils.
All the main leading non-axisymmetric terms\change{}{, except $B_{0,1}$, }have been reduced\change{, especially the $B_{1,2}$ term}.
It indicates that the optimized coils have a better quasi-axisymmetric magnetic field across the whole plasma, although we are \changenew{only}{} trying to improve QA quality on \changenew{}{only} one flux surface.
The optimized coils are shown in \Fig{QA_opt_coils}.
Compared to the original coils, the optimized ones do not have unrealistic increase in geometry complexity (just naively comparing by eye).
Note that this is not a comprehensive optimization and \change{}{the only objective function, $F_{QS}$ is not at a local minimum after 10 iterations.}
We are also not attempting to compare which coil set is better.
The results demonstrate that the QA quality could be directly optimized during coil designs.
In future studies, we could actually employ other powerful optimization algorithms, like the modified Newton method (MN) \cite{Zhu_2018_Newton} since the gradient and Hessian have been analytically calculated.

\section{Conclusions} \label{discussion}
The recently developed Hessian matrix method for analyzing error field sensitivity has been extended to identify \change{dangerous}{important} coil deviations that will open magnetic islands or degrade quasi-symmetry.
To manipulate islands, a new functional that evaluates resonant perturbations is incorporated.
The quasi-symmetry quality on one flux surface is assessed by computing Fourier components in Boozer coordinates.
With the two new figures of merit for quantifying error fields, we have applied the Hessian matrix method to the CFQS device which is under construction.
We have successfully found the most effective coil perturbation scheme that will enlarge the 4/11 islands, as shown in \Fig{ev1_rp_pp}.
In addition, the \change{dangerous}{important} coil perturbations deteriorating the quasi-axisymmetry on the half-toroidal-flux surface are also identified (\Fig{ev_qa_coils}).
These results will provide insights for the upcoming coil manufacture and assembly: avoiding \change{dangerous}{critical} coil deformations and relaxing coil tolerance at insensitive parts.
A simple estimation indicates that the maximum allowable coil deviation is about 5 mm if 10\% of QA quality deterioration is acceptable.
We have also demonstrated the ability \changenew{of improving}{to improve} coil designs towards better QA quality.
This could be used in future coil optimization studies.

In this paper, each coil is considered independently, but the leading eigenvectors shown above are surprisingly preserving stellarator symmetry and periodicity.
This is not true for all the eigenvectors and some eigenvectors with smaller eigenvalues do not retain stellarator symmetry or periodicity.
There is a possibility that the eigenvectors preserving stellarator symmetry (and/or periodicity) will be more effective than others if the figure of merit and present coils are both symmetric (and/or periodic).
Rigorous mathematical explorations are remaining for future work.
Here, we only employ Fourier representation to describe coil geometries.
In general, any other representation could be used.
Some representation might be able to resolve the discrepancy between the magnitude in parameter space and in real space as displayed in Sec. \ref{result_tol}.
So far we have only considered coil filaments.
When coils are not too close to the plasma, this approximation is sufficient, as the magnetic field produced by coils is proportional to $r^{-3}$.
However, coils are eventually built with finite thickness.
A finite-build model would be still essential.
We can use the Hessian matrix method as a pre-processing step since it is fast and the sensitivity information could provide valuable guidance for detailed engineering analyses.
To analytically calculate the derivatives of resonant perturbations and quasi-symmetry, we assume the flux surface \change{}{and the straight field-line coordinates system} do not move under \emph{small} perturbations.
\change{Although flux surfaces might actually move/deform under perturbations, t}{T}he fact that we successfully manipulated the island size (\Fig{ev1_rp_pp}) proves this simplification is valid here.
Additionally, as shown in Sec. \ref{QA_opt}, the global QA quality was improved by minimizing $F_{QS}$ in \Eqn{fqs} using the same assumption.
\change{Of course, by using finite difference to calculate the Hessian matrix, there would be no necessity to keep this assumption and all the calculations would be self-consistent.}
{This linear approximation is not essential and might not be as accurate as nonlinear analysis since in reality flux surfaces and coordinates systems are likely changing under coil perturbations.
By using the finite difference to calculate the Hessian matrix, the assumption can be relaxed and all the calculations would be self-consistent, although the finite difference suffers numerical noises and takes much more time.}
\changenew{}{Careful validations on the linear approximation should be carried out in the future work.}
While we have \changenew{only} discussed using the Hessian matrix method for \changenew{}{only} studying error fields from stellarator coils in vacuum configurations, it could \changenew{}{also} be \changenew{also}{} applied to equilibria with finite plasma currents, or even in tokamaks and other general fields to obtain sensitivity information.



\section*{Acknowledgments}
The authors gratefully appreciate fruitful discussions with A. Brooks, Y. Suzuki, M. Landreman S. Lazerson and N. Pomphrey.
This work was supported by the U.S. Department of Energy under Contract No. DE-AC02-09CH11466 through the Princeton Plasma Physics Laboratory and the Max-Planck-Princeton Center for Fusion and Astro. Plasma Physics.HL and YX would like to acknowledge the National Science Foundation of China under Grant No. 11820101004 and the National Key R\&D Program of China under Grant No. 2017YFE0301705.

\begin{appendices}
\numberwithin{equation}{section}

\section{Optimal Fourier resolution for CFQS modular coils} \label{nf_scan}
\begin{figure}[ht]
    \centering
    \includegraphics[width=0.75\textwidth]{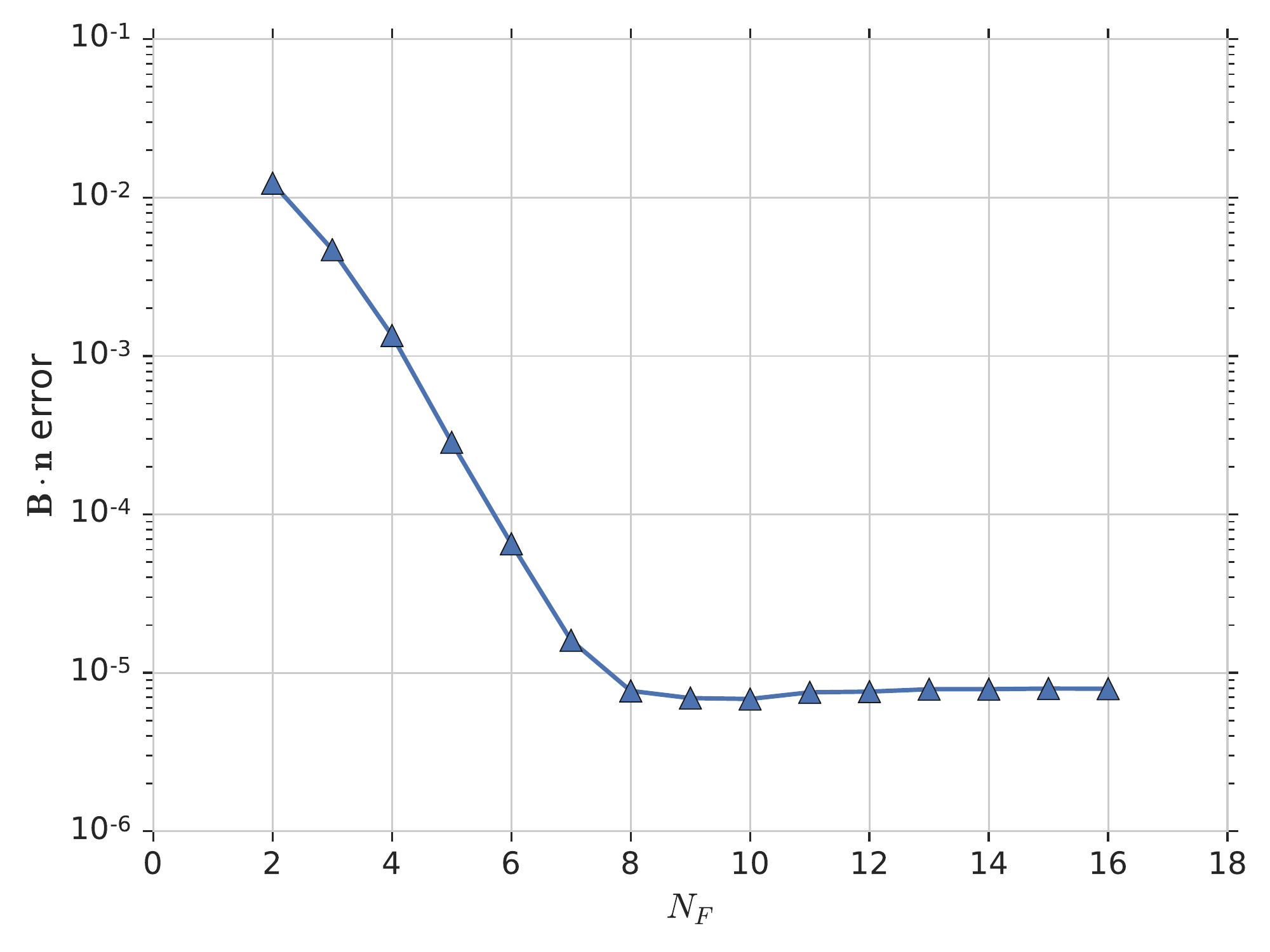}
    \caption{Residual normal field errors produced by CFQS coils fitted with different numbers of Fourier coefficients.}
    \label{fig:scan_nf}
\end{figure}

Each CFQS modular coil is piece-wisely described by 48 points in space.
However, FOCUS employs Fourier representation.
We need to fit CFQS coil data with Fourier coefficients.
First of all, we used cubic spline interpolation to smooth coils and each coil now has 256 points.
Afterwards, FOCUS read coil data, performed Fourier transformations and calculated the residual $\vect{B} \cdot \vect{n}$ error on the last closed flux surface (obtained from free-boundary VMEC runs).
\Fig{scan_nf} shows how the residual normal field error varies when we increase the resolution of Fourier coefficients.
$N_F$ is the maximum Fourier mode truncated in each coil.
The residual error is exponentially reduced with increasing $N_F$.\change{ and it converges after $N_F>8$.
In consideration of `safety factor', $N_F = 10$ was selected (besides, $N_F = 10$ has the minimum error).}
{Since $N_F = 10$ has the minimum error, we will use $N_F = 10$ for future calculations.}
Then the total number of degrees of freedom, as listed in \Eqn{fourier}, is $1024=16 \times (6 \times 10 +3+1)$.

Actually, the number of required Fourier coefficients to adequately describe CFQS coils should be smaller than 1024. 
The spectrum of Fourier modes can be condensed by carefully choosing the parameterization angle.
Besides, some type of coil with simpler geometry might need less Fourier coefficients.
But we are not going to explore these issues in this paper.

\section{Quadratic approximation of QA quality when the gradient is non-zero} \label{quadratic}
In Sec. \ref{result_qa}, the Hessian matrix for QA quality on the half-toroidal-flux surface is not positive definite.
It implies the figure of merit is not at a strict local minimum with CFQS coils.
However, the quadratic approximation assumes the gradient is negligible.
Therefore, we need to test if \Eqn{quadratic} is still valid.

\begin{figure}[ht]
    \centering
    \includegraphics[width=0.75\textwidth]{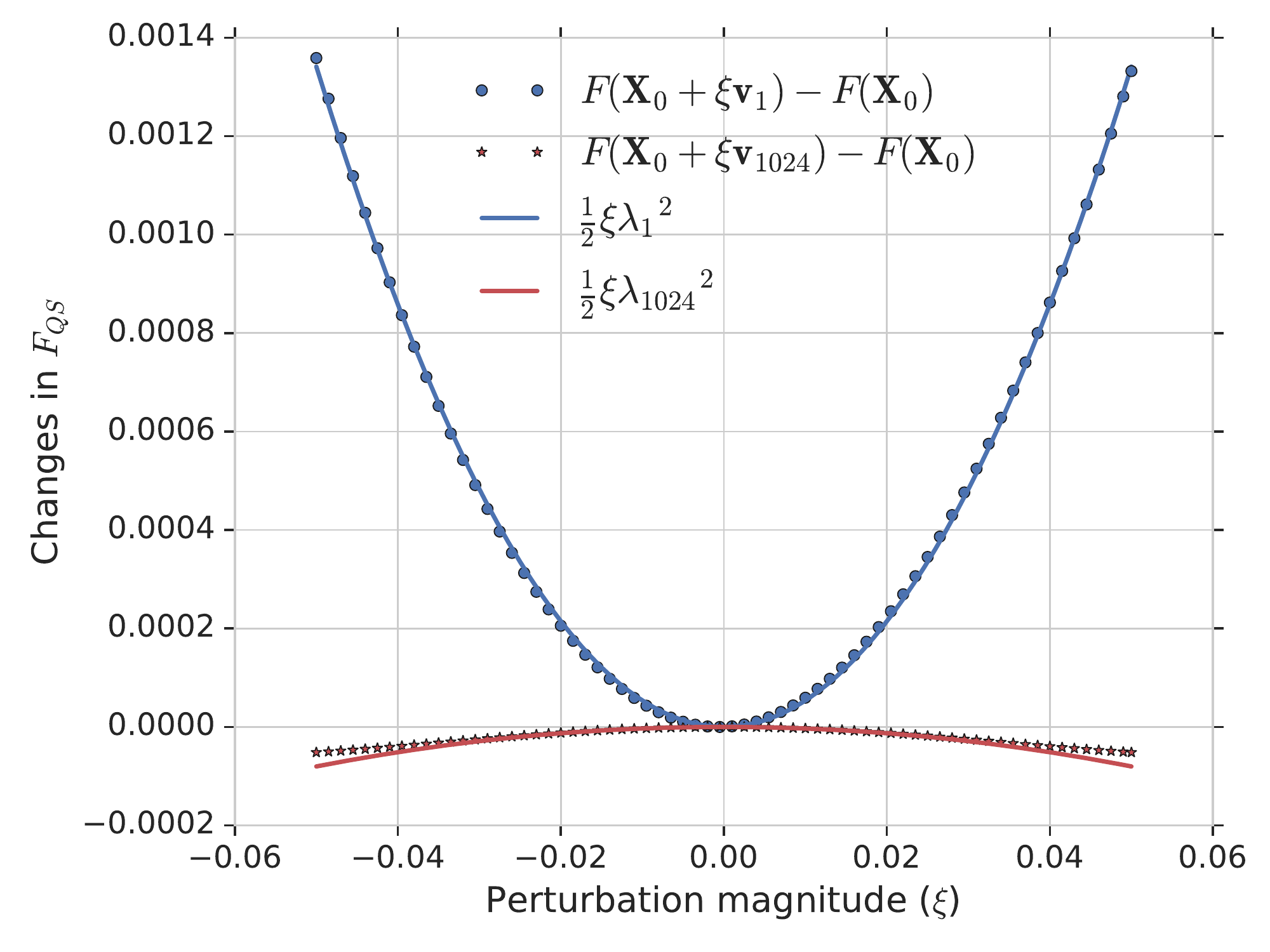}
    \caption{Changes in the QA quality when perturbing coils in directions of the first and last eigenvectors. $\xi < 0$ means perturbing coils in opposite directions.}
    \label{fig:quadratic_qa}
\end{figure}

Choose a perturbation direction, \eg the first principal eigenvector of the Hessian matrix, apply different magnitudes of perturbation and then use FOCUS to compute the QA quality with each perturbed coil set.
\Fig{quadratic_qa} shows the relationship between changes in the QA quality $\Delta F_{QS}$ and the perturbation magnitudes $\xi$ in both the first principal eigenvector $\vect{v}_1$ and the last principal eigenvector $\vect{v}_{1024}$.
The two lines are consistent with the predicted quadratic lines, even in a relatively sizable range.
The results imply that the QA quality is marginally close to a local minimum and the quadratic approximation is still valid.

\end{appendices}

\section*{References}
\bibliographystyle{unsrt}
\bibliography{sensitivity}

\end{document}